\documentclass[12pt]{article}
\usepackage{amsfonts, amsmath, amssymb, cite, graphicx, }

\usepackage{color}
\usepackage[all, knot]{xy}
\usepackage{tikz}
\usepackage{subfigure}
\usepackage{braket}

\usepackage[utf8]{inputenc}
\usepackage{ulem}

\usepackage{bm}
\usepackage{epstopdf}
\usepackage[footnotesize]{caption}
\usepackage{amsthm}
\usepackage{amsthm,amsmath,amssymb}
\usepackage{enumitem}
\usepackage{mathrsfs}
\usepackage{makecell}
\usepackage{diagbox}

\usepackage[margin=3cm]{geometry}

\begin{document}

\begin{titlepage}
%\begin{flushright}
%TIT/HEP-6XX \\
%mm,  2016
%\end{flushright}
\vspace{0.5cm}
\begin{center}
{\Large \bf AdS$_{3}$ spacetime wormhole fluctuations and their impact on false vacuum decay}

\lineskip .75em
\vskip 2.5cm
{\large Hong Wang$^{a}$ and Jin Wang$^{b,}$\footnote{Corresponding author, jin.wang.1@stonybrook.edu} }
\vskip 2.5em
 {\normalsize\it $^{a}$State Key Laboratory of Electroanalytical Chemistry, Changchun Institute of Applied Chemistry, Chinese Academy of Sciences, Changchun 130022, China\\
 $^{b}$Department of Chemistry and Department of Physics and Astronomy, State University of New York at Stony Brook, NY 11794, USA}
\vskip 3.0em
\end{center}
\begin{abstract}
In this work, we studied the boundary wormhole fluctuations of AdS$_{3}$ spacetime and their impact on false vacuum decay. The model consists of three dimensional
general relativity coupled to a real scalar field with self-interaction potential. Our analysis focuses on handlebody spacetimes, which correspond to boundary wormhole fluctuations. These fluctuations result in variations in the spacetime genus. Using the dilute wormhole gas approximation, we computed the boundary wormhole creation rate in the semiclassical region. Our results indicate that boundary wormholes are more likely to be created in spacetimes with larger absolute values of the cosmological constant. Furthermore, we find that smaller wormholes are more easily created compared to larger ones. Additionally, we investigated the impact of boundary wormhole fluctuations on false vacuum decay. The vacuum decay rate increases when the effects of wormhole fluctuations are considered.  We show that all results remain qualitatively unchanged when certain parameter values are varied.
\end{abstract}
\end{titlepage}

\baselineskip=0.7cm

\tableofcontents
\newpage
\section{Introduction}
\label{sec:1}
The local properties of classical spacetime are determined by the metric, which can be obtained by solving the Einstein equations.  In addition to local properties, spacetime may possess non-trivial topological invariants, such as genus and homology groups~\cite{MJ}.  These topological invariants are global properties of spacetime and are independent of the metric~\cite{EW,DM1230}. This independence implies that the Einstein equations do not determine the topological invariants.  Moreover, the topological invariants of spacetime may correspond to observable effects~\cite{SW, MJ, AB}.  It is widely believed that quantum fluctuations can change the topological invariants of spacetime~\cite{EW, SR, SS}. Thus, topological invariants are important topics in both classical and quantum gravity.

The renormalization problem of four dimensional quantum gravity remains unresolved~\cite{BB}. However, three dimensional quantum gravity is renormalizable and lacks bulk graviton excitations~\cite{EW2}. It has been shown that three dimensional Einstein pure gravity (with no matter except dark energy) is equivalent to the Chern-Simons topological field theory~\cite{EW2}. Thus, three dimensions provide an ideal framework for studying topological issues in quantum gravity. Despite our universe being at least macroscopically four dimensional, insights from three dimensional studies are believed to offer valuable perspectives~\cite{SS,EW2}.

When the cosmological constant is positive or zero, there is a finite class of topologically non-equivalent three dimensional  spacetimes. In contrast, with a negative cosmological constant, there is an infinite class of topologically non-equivalent three dimensional spacetimes~\cite{SC}. Thus, AdS$_{3}$ spacetime exhibits rich topological properties. The conformal boundary of the genus zero Euclidean AdS$_{3}$ spacetime is a cylinder. All genus one AdS$_{3}$ spacetimes are handlebodies, whereas higher genus AdS$_{3}$ spacetimes can be either handlebodies or non-handlebodies~\cite{XY1, XY2, HSB}.
A handlebody has only one boundary, while a non-handlebody has multiple disconnected boundaries~\cite{XY1,XY2,HSB,JL}. That is, in a non-handlebody spacetime, the boundaries are connected only through the bulk and cannot be connected through the boundary itself~\cite{JL}. Studies by Maldacena, Maoz, and other researchers  have highlighted certain puzzles associated with the non-handlebody solutions of the Einstein equations in the context of AdS/CFT duality~\cite{JL,PE,PEO,MVR,DMJ}. However, in three dimensional spacetime,  some research suggests that the contribution of non-handlebody spacetimes to the partition function is small compared to that of  handlebody spacetimes~\cite{XY2, HSB}. Therefore, for simplicity, this work focuses on AdS$_{3}$ handlebody spacetimes.

Significant theoretical progress has been achieved in studying spacetimes with non-zero genus.  In 1989, Witten demonstrated that, when the cosmological constant is zero, the partition function of three-dimensional pure quantum gravity can be expressed as the integral of the Ray-Singer analytic torsion (a topological invariant) over the moduli space of the Einstein equations~\cite{EW}. In 1993, Carlip introduced a method to compute the partition function of three dimensional quantum gravity, accounting for wormhole fluctuations in the limit of a small cosmological constant. He showed that the partition function exhibits  divergence~\cite{SC2}.
In 2000, Krasnov showed that the regularized Einstein-Hilbert action of the AdS$_{3}$ handlebody spacetime is equivalent to the Takhtajan-Zograf action, which corresponds to the action of the Liouville field defined on the two dimensional boundary of the bulk AdS$_{3}$ spacetime~\cite{KK}. Subsequently, Takhtajan, Teo, and Park proved that this equivalence holds even when the boundary surface contains elliptic or parabolic singularities (cone points or punctures, respectively)~\cite{TT,PTT,PT}.  These findings strongly support the AdS$_{3}$/CFT$_{2}$ duality. In recent years, researchers have employed new methods to study the partition function of thermal AdS$_{3}$ spacetime, incorporating  contributions from boundary graviton excitations~\cite{AE,NAS}. Other significant studies have also been undertaken; see~\cite{SCO,SC3,KB,AE2,JKA,SA,H,PSD,SH1,SH2,SH3,SH4} and references therein. We will not enumerate them all here.

Despite significant progress in three dimensional topological quantum gravity, the challenge of wormhole fluctuations remains unresolved~\cite{EW,SC2}. In this work, we aim to study the boundary wormhole fluctuations of AdS$_{3}$ spacetime and their influence on false vacuum decay. All spacetimes considered are AdS$_{3}$ handlebody solutions of the Euclidean Einstein equations, which correspond to the boundary wormhole fluctuations. The local properties of various AdS$_{3}$ handlebodies are identical, as they are determined by the Euclidean Einstein equations, while their global structures differ~\cite{KK}. The model consists of a real scalar field minimally coupled to three dimensional general relativity. The scalar potential has two minimal values, and the difference between these values is assumed to be small, ensuring the validity of the thin wall approximation.  When the scalar field occupies one of the potential minima, it acts as a cosmological constant.  This model describes both wormhole fluctuations and false vacuum decay, with no restriction on the cosmological constant being close to zero.

The Euclidean AdS$_{3}$ spacetime action is divergent and requires regularization~\cite{KK}. The process of boundary wormhole fluctuation is accompanied by changes in the spacetime genus. We work in the semiclassical region and use the Euclidean path integral approach to compute the boundary wormhole creation rate. Using the dilute wormhole gas approximation, we transform the tunneling action from the Schottky space to the Siegel upper half space. This approximation simplifies the integral domain, which is a subspace  of the fundamental domain of the modular group Sp$(2g,\mathbb{Z})$ in the Siegel upper half space. Consequently, integration over the physically non-equivalent AdS$_{3}$ spacetimes becomes feasible.

Our results indicate that boundary wormholes are more easily created in AdS$_{3}$ spacetime with a more negative cosmological constant.  We find that small boundary wormhole are more likely to occur than larger ones. This suggests that large scale wormholes are difficult to create.  We also studied the influence of boundary wormhole fluctuations on AdS$_{3}$ vacuum decay. Both the dilute instanton gas approximation and the dilute wormhole gas approximation were used.
Our analysis shows that including boundary wormhole fluctuations increases the vacuum decay rate. We also observe that defining the initial state within our framework breaks modular symmetry. Although there are two unspecified parameters in our model, we demonstrate that their values do not qualitatively affect the results.  Throughout this study, we work in units where $8\pi G=\hbar=k_{B}=c=1$.

\section{Boundary wormhole fluctuations and false vacuum decay}
\label{sec:3.0}

A simple model for studying quantum gravity consists of three dimensional general relativity coupled to a real scalar field with a self-interaction potential. The Euclidean action for this model is given by~\cite{JJ}
\begin{equation}
\label{eq:3.1}
S_{E}=\int dx^{3}\sqrt{g}\{-\frac{1}{2}R+\frac{1}{2}g^{\mu\nu}\partial_{\mu}\phi\partial_{\nu}\phi+V(\phi)\} +S_{GHY}.
\end{equation}
Here, $S_{E}$ represents the Euclidean action of the total system.  $R$  and $S_{GHY}$ represent the Ricci scalar  and the Gibbons-Hawking-York surface term, respectively. $\phi$ represents the scalar field, and $V(\phi)$ represents the scalar potential. In quantum physics, both real time and imaginary time coordinates are commonly used. In this work, we choose to  use the imaginary time coordinates. In this coordinate system, the metric signature  of the three dimensional spacetime is $(+,+,+)$, and the  volume element is $\sqrt{g}dx^{3}$. (If real time coordinates are used instead, the metric signature becomes $(-,+,+)$ and $\sqrt{g}$ should be replaced with $\sqrt{-g}$.)

\begin{figure}[tbp]
\centering
\includegraphics[width=5cm]{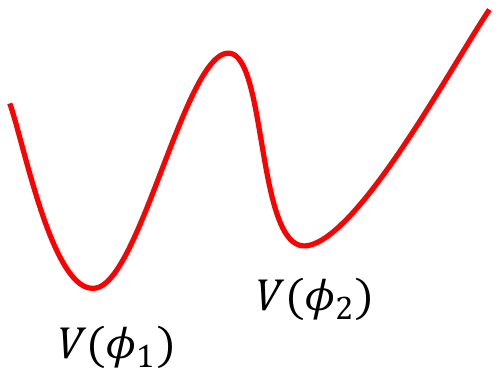}
\caption{\label{fig:3} Diagram of the scalar potential. It has two minimal values, denoted as  $V(\phi_{1})$ and $V(\phi_{2})$.  }
\end{figure}

In this work, we focus on the case where the scalar potential has two minimal values,  as shown in figure~\ref{fig:3}. These minimal values are labeled as $V(\phi_{1})$ and $V(\phi_{2})$, with $V(\phi_{1})<V(\phi_{2})<0$.  Usually, the states $V(\phi_{1})$ and $V(\phi_{2})$ are called the true vacuum state and the false vacuum state, respectively~\cite{SF}.  For simplicity, we assume $\epsilon=V(\phi_{2})-V(\phi_{1})$ to be a small quantity. In this case,  the thin wall approximation is valid~\cite{SF,HJ}. Extending this model to scenarios where the potential has more local minimal values  is straightforward.

When the scalar field is in the vacuum state $V(\phi_{1})$ (or $V(\phi_{2})$), it remains constant over time in classical dynamics, and its kinetic energy vanishes~\cite{JM73}. Consequently, the Euclidean action of the scalar field in this case reduces to $\int dx^{3}\sqrt{g}V(\phi_{1})$ (or $\int dx^{3}\sqrt{g}V(\phi_{2})$). Moreover, the cosmological constant $\Lambda$ corresponds to the Euclidean action $\int dx^{3}\sqrt{g}\Lambda$. When the scalar field remains unchanged over time, it is reasonable to set  $V(\phi_{1})$ (or $V(\phi_{2})$) equal to $\Lambda$, in which case the action $\int dx^{3}\sqrt{g}V(\phi_{1})$ (or $\int dx^{3}\sqrt{g}V(\phi_{2})$) becomes $\int dx^{3}\sqrt{g}\Lambda$. This indicates that when the scalar field is in the vacuum state $V(\phi_{1})$ (or $V(\phi_{2})$), it is effectively equivalent to a cosmological constant. This equivalence can also be demonstrated by comparing the stress energy tensor of the scalar field with that of the cosmological constant~\cite{JM73}.  In addition, the Friedmann equations indicate that the cosmological constant can drive inflation. Therefore, the scalar field in its (false) vacuum state is often used to study the inflationary phase of the early universe~\cite{DB73}.

The action \eqref{eq:3.1} is generally valid regardless of the spacetime genus. The model defined by this action  can describe both wormhole fluctuations and false vacuum decay. When the scalar field resides at one of the potential minima, it effectively serves as a cosmological constant, allowing the model to describe wormhole fluctuations. In another situation, if quantum fluctuations of the spacetime topological structure are neglected, the model \eqref{eq:3.1} can be used to study false vacuum decay.
In this work, we consider both wormhole fluctuations and false vacuum decay. Thus, it is necessary to include a scalar field in the model. The false (true) vacuum state $V(\phi_{2})$ ($V(\phi_{1})$ ) of the scalar field is equivalent to a cosmological constant. Thus, we may refer to $V(\phi_{1})$ or $V(\phi_{2})$ as a cosmological constant to emphasize its relationship with wormhole fluctuations.  At other times, we refer to $V(\phi_{1})$  and $V(\phi_{2})$  as the true vacuum state and false vacuum state, respectively, to highlight the vacuum decay dynamics. Additionally,  in some equations, we may use the symbol $\Lambda$ to represent the cosmological constant for convenience. This flexibility  in terminology are necessary to clearly describe different physical processes and to avoid confusion.

When the scalar field is in the state $V(\phi_{1})$ or $V(\phi_{2})$, the Einstein equations reduce to
\begin{equation}
\label{eq:m3.1}
R=6\Lambda.
\end{equation}
In our model, the value of the cosmological constant $\Lambda$ equals $V(\phi_{1})$ ($V(\phi_{2})$) when the scalar field is in the state $V(\phi_{1})$ ($V(\phi_{2})$). Equation \eqref{eq:m3.1} admits infinitely many solutions~\cite{XY1,KK,SS,MN}, all of which are AdS$_{3}$ spacetimes.  These solutions can be categorized into handlebodies and non-handlebodies~\cite{XY1, XY2, HSB}.  Some studies indicate that the quantum effects of handlebodies are dominant~\cite{XY2, HSB}. Thus, for simplicity, we focus on AdS$_{3}$ handlebody solutions.

\begin{figure}[tbp]
\centering
\includegraphics[width=10cm]{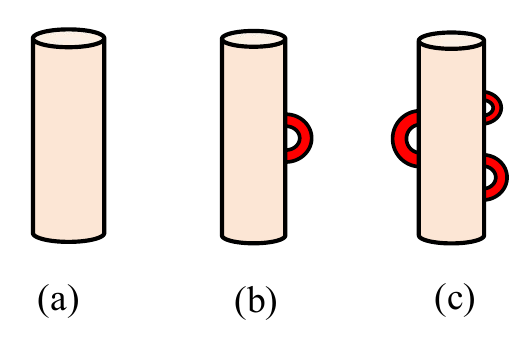}
\caption{\label{fig:m1} Diagram of various AdS$_{3}$ handlebodies. Figures (a), (b) and (c) represent the AdS$_{3}$ handlebody with genus zero, one and three, respectively. The cylinder represents the parent universe. The red strips  represent the boundary wormholes. }
\end{figure}

Although all solutions to equation \eqref{eq:m3.1} have the same local properties, their global properties differ~\cite{KK,SS,MN}. In other words, different AdS$_{3}$ spacetimes may possess  distinct topological invariants~\cite{KK,SS,MN}, which are not determined by equation \eqref{eq:m3.1}.  Locally, every point in any AdS$_{3}$ handlebody has the same Ricci scalar.  Globally, however,  various AdS$_{3}$ spacetimes may have different genus~\cite{KK,SS,MN}.  Figure~\ref{fig:m1} illustrates some of these solutions.  The cylinder shown in figure~\ref{fig:m1}(a) represents a genus zero AdS$_{3}$ spacetime. We will refer to it as the parent universe. Figures~\ref{fig:m1}(a), ~\ref{fig:m1}(b) and ~\ref{fig:m1}(c) depict  AdS$_{3}$ handlebodies of genus zero,  one and three, respectively. The red strips in figure~\ref{fig:m1} represent boundary wormholes, which can be produced through quantum tunneling~\cite{SR}. We refer to these processes as boundary wormhole fluctuations (BWFs). The handlebodies shown in figures~\ref{fig:m1}(b) and~\ref{fig:m1}(c) can be interpreted as processes through which boundary wormholes are created in the parent universe. Figure~\ref{fig:m1} clearly shows that BWFs alter the spacetime genus.

Locally, any AdS$_{3}$ handlebody is homogenous and isotropic. Thus, one can use the metric~\cite{JJ}
\begin{equation}
\label{eq:3.2}
ds^{2}=d\tau_{E}^{2}+\rho(\tau_{E})^{2}(d\theta^{2}+sin^{2}\theta d\varphi^{2}).
\end{equation}
to describe the local properties of various AdS$_{3}$ handlebodies.  Here, $\tau_{E}$ and $\rho(\tau_{E})$ represent the Euclidean proper time and the scale factor, respectively. The parameters $\varphi$ and $\theta$ range as $0\leq\varphi\leq2\pi$ and $0\leq\theta\leq\pi$, respectively. Given any point $P$ in an AdS$_{3}$ handlebody, there exists a neighborhood of $P$ that is spherically symmetric. Within this neighborhood, the metric \eqref{eq:3.2} is well defined.    In the special case of a genus zero AdS$_{3}$ spacetime (the parent universe), the spacetime is globally homogeneous and isotropic, not just locally. Hence, the metric \eqref{eq:3.2} is valid throughout the entire parent universe, and all properties of the parent universe can be described by it.   However, higher genus AdS$_{3}$ handlebodies lack global spherical symmetry. Therefore, this metric cannot be globally applied to higher genus AdS$_{3}$ handlebodies.

Now consider false vacuum decay in the parent universe.  In the semiclassical region, the Coleman-De Luccia theory gives the vacuum decay rate as~\cite{SF}
\begin{equation}
\label{eq:3.8}
\Gamma\approx\mathrm{exp}\big\{-\big(S_{E}(\phi_{bounce})-S_{E}(\phi_{2})\big)\big\}\equiv e^{-B}.
\end{equation}
Here, $\phi_{bounce}$ represents the bounce solution of the scalar field. We refer to $S_{E}(\phi_{2})$ as the background term,  and $B$ as the total tunneling action. In deriving equation \eqref{eq:3.8}, Both the thin wall approximation and the dilute instanton gas approximation are used. The bounce solution can  be visualized as  a three dimensional spherically symmetric bubble (the bounce bubble) embedded in a sea of false vacuum~\cite{SF}. The radius of the bounce bubble $\bar{\rho}$ is determined by the condition $\partial B/\partial \bar{\rho}=0$~\cite{SF}. Thus, the tunneling action can be written as $B=B_{in}+B_{wall}+B_{out}$, where $B_{in}$ and $B_{out}$ denote the tunneling actions inside and outside the bubble, respectively, and $B_{wall}$ represents the tunneling action on the domain wall.

\begin{figure}[tbp]
\centering
\includegraphics[width=9cm]{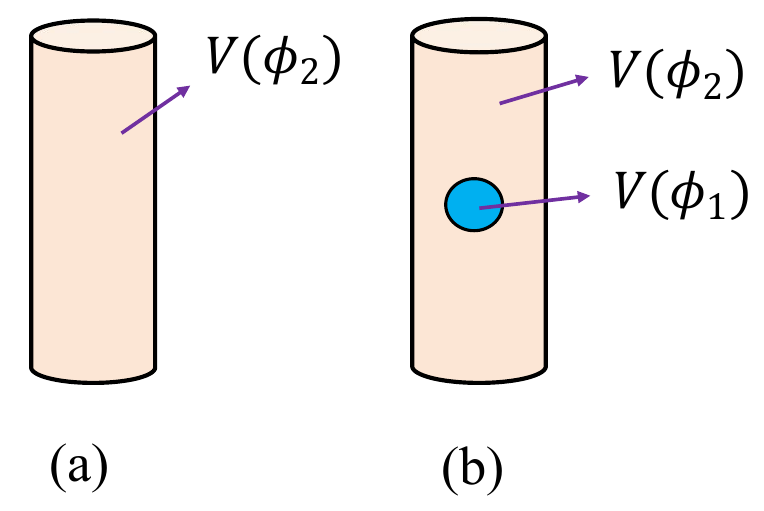}
\caption{\label{fig:m2} Diagram of false vacuum decay. Figures (a) and (b) represent the background term and the bounce solution, respectively. The blue ball in figure (b) represents the bubble. }
\end{figure}

The background term and the bounce solution can be represented by figures~\ref{fig:m2}(a) and ~\ref{fig:m2}(b), respectively.  The blue ball in figure~\ref{fig:m2}(b) represents the bubble. Inside and outside the blue ball, the cosmological constant takes the values $V(\phi_{1})$ and $V(\phi_{2})$, respectively. In figure~\ref{fig:m2}(a), the scalar field is entirely in the state $V(\phi_{2})$. Figure~\ref{fig:m2}(b) describes the process of false vacuum decay. Comparing figures~\ref{fig:m2}(a) and ~\ref{fig:m2}(b), one can see that false vacuum decay can also be interpreted as the creation of a bubble. Consequently, the vacuum decay rate can be understood as the bubble creation rate.

\begin{figure}[tbp]
\centering
\includegraphics[width=10cm]{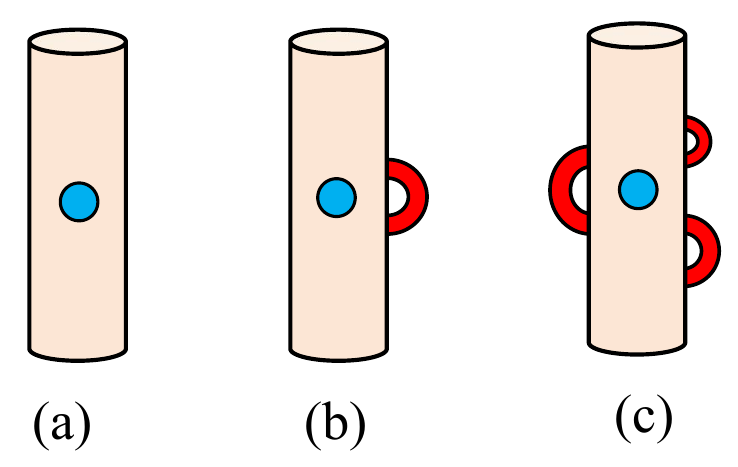}
\caption{\label{fig:m3} Diagram of false vacuum decay in various AdS$_{3}$ spacetimes. The cylinders represent the parent universe, while the blue balls and red strips represent the bubble and boundary wormholes, respectively. }
\end{figure}

The dynamical equation of the scalar field in various AdS$_{3}$ spacetimes is the same. It is the Klein-Gordon equation. This implies that false vacuum decay  can occur not only in the genus zero AdS$_{3}$ spacetime but also in higher genus AdS$_{3}$ spacetimes. In other words, bubble creation is possible in both genus zero and higher genus AdS$_{3}$ spacetimes, as illustrated in figure~\ref{fig:m3}. To calculate the bubble creation rate (i.e., false vacuum decay rate), all possible modes of bubble creation must be considered. It is important to note that different modes of bubble creation correspond to different AdS$_{3}$ handlebodies, and these handlebodies can be interpreted as representing different BWFs.  Therefore, summing over all possible modes of bubble creation is equivalent to summing over all BWFs. As a result, the total bubble creation rate ( or total false vacuum decay rate) can be expressed as:
\begin{equation}
\label{eq:3.2m}
\Gamma_{tot}=\sum_{\mathcal{M}}e^{-B}.
\end{equation}
Here, $\sum_{\mathcal{M}}$ represents the summation over all physically non-equivalent BWFs. Various BWFs correspond to different handlebodies. The global properties of these handlebodies differ, while their local properties (such as the metric) remain the same. Thus, the summation in equation \eqref{eq:3.2m} does not include a summation over metrics.

In sections~\ref{sec:3} and ~\ref{sec:4}, we will study the total tunneling action $B$ and the summation $\sum_{\mathcal{M}}$  in equation \eqref{eq:3.2m} in detail. We will demonstrate that the total tunneling action $B$ varies depending on the specific mode of false vacuum decay. Obviously,  if the effects of BWFs are neglected, equation \eqref{eq:3.2m} reduces to equation \eqref{eq:3.8}.
Additionally, in section~\ref{sec:5}, we will show that once the total tunneling action $B$ and the total tunneling rate $\Gamma_{tot}$  for false vacuum decay are determined, various interesting properties of AdS$_{3}$ BWFs can be derived from these quantities. Therefore, deriving the total tunneling action $B$ and the total tunneling rate $\Gamma_{tot}$  are the main tasks of this study.

The conventional Coleman-De Luccia theory (equation \eqref{eq:3.8}) describes false vacuum decay at zero temperature~\cite{SF,OJ74}. We attempt to generalize this theory to include the effects of wormhole fluctuations, as shown in equation \eqref{eq:3.2m}. Equation \eqref{eq:3.2m} reduces to equation \eqref{eq:3.8} in the limit where the effects of wormhole fluctuations vanish. Thus, both equations \eqref{eq:3.8} and \eqref{eq:3.2m} are valid in the  zero temperature regime. In the case of non-zero temperature, the scalar field may be excited to highly excited states, and the Coleman-De Luccia theory should be replaced by other theories\cite{OJ74,AD741,AD742}. These effects introduce additional complexity into the model. Therefore, for simplicity, we focus on the zero temperature case in this study.

\section{The tunneling action}
\label{sec:3}

\subsection{The total tunneling action}
\label{sec:3.1}

\begin{figure}[tbp]
\centering
\includegraphics[width=4cm]{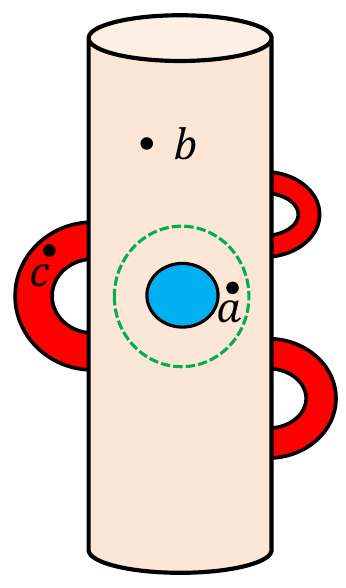}
\caption{\label{fig:m4} Diagram of false vacuum decay in a high genus AdS$_{3}$ handlebody. The cylinder represents the parent universe. The blue ball and the red strips represent the bubble and the boundary wormholes, respectively. The green circle represents a spherical surface that is larger than the brane of the bounce bubble. }
\end{figure}

In this section, we focus on calculating the total tunneling action $B$ in high genus spacetimes. Although high genus spacetimes do not possess global spherical symmetry, one can always find a locally spherically symmetric region, as illustrated in figure~\ref{fig:m4}. The bounce bubble (with radius denoted as $\overline{\rho}$) resides within the parent universe. Thus, a spherically symmetric region containing the bounce bubble can always be found, such as the region inside the green circle (representing a spherical surface with radius larger than $\overline{\rho}$ ) in figure~\ref{fig:m4}. We denote the radius of this region as $\rho_{s}$ ($\rho_{s}>\overline{\rho}$ ). The exact value of $\rho_{s}$ is not important.

The region with radius smaller than $\rho_{s}$ in figure~\ref{fig:m4} is chosen to have spherical symmetry. Thus, within this region, the metric \eqref{eq:3.2} is well defined. Consequently, the Ricci scalar can be expressed as~\cite{JJ}
\begin{equation}
\label{eq:3.3}
R=-4\frac{\ddot{\rho}}{\rho}+\frac{2}{\rho^{2}}(1-\dot{\rho}^{2}),
\end{equation}
where, $\dot{\rho}\equiv d\rho/d\tau_{E}$. Substituting equations \eqref{eq:3.2} and \eqref{eq:3.3} into equation \eqref{eq:3.1},  the Euclidean action \eqref{eq:3.1} in the region $\rho\leq\rho_{s}$ can be written as
\begin{equation}
\label{eq:3.4}
S_{E}(\rho\leq\rho_{s})=\int d\theta d\varphi\int_{0}^{\rho_{s}}d\tau_{E}\sqrt{g}\big\{-\frac{\dot{\rho}^{2}}{\rho^2}-\frac{1}{\rho^2}+\frac{1}{2}\dot{\phi}^{2}+V(\phi)\big\}.
\end{equation}
Here, $S_{E}(\rho\leq\rho_{s})$ represents the Euclidean action for the region $\rho\leq\rho_{s}$ of the model defined by equation \eqref{eq:3.1}.

From equations \eqref{eq:3.1} to \eqref{eq:3.4}, we utilize the fact that  the Gibbons-Hawking-York surface term $S_{GHY}$ is canceled out by the boundary term generated in the partial integral with respect to the Ricci scalar $R$. The Euclidean action plays a central role in the Euclidean path integral. In the semiclassical region, this method is widely used to study quantum gravity with a negative cosmological constant~\cite{SR,SAX,XY1,SC2,AE,NAS}. Thus, we will use this method to study the model defined by equation \eqref{eq:3.1}.

From equation \eqref{eq:3.4},  the total Hamiltonian density ($\mathscr{H}_{tot}$) of the system in the region $\rho\leq\rho_{s}$ is given by
\begin{equation}
\label{eq:3.4a}
\mathscr{H}_{tot}=\sqrt{g}\big\{-\frac{\dot{\rho}^{2}}{\rho^2}+\frac{1}{\rho^2}+\frac{1}{2}\dot{\phi}^{2}-V(\phi)\big\}.
\end{equation}
General covariance requires the total Hamiltonian density to be zero~\cite{HJ,CK,HJ2,HJ3}. Using  $\delta S_{E}/\delta\phi$ =0 and $\mathscr{H}_{tot}=0$, one can obtain the dynamical equations~\cite{JJ}
\begin{equation}
\label{eq:3.5}
\ddot{\phi}=\frac{\partial V}{\partial \phi},
\end{equation}
\begin{equation}
\label{eq:3.6}
\dot{\rho}^{2}=1+\rho^{2}\big(\frac{1}{2}\dot{\phi}^{2}-V(\phi)\big).
\end{equation}
Equation \eqref{eq:3.5} is the Klein-Gordon equation, while equation \eqref{eq:3.6} is the Friedmann equation. Note that in equation \eqref{eq:3.5}, the Hubble drag term $2\dot{\rho}\dot{\phi}/\rho$ has been neglected, as it is small under the thin wall approximation~\cite{SF}.

Substituting equation \eqref{eq:3.6} into equation \eqref{eq:3.4}, the Euclidean action \eqref{eq:3.4} simplifies to
\begin{eqnarray}\begin{split}
\label{eq:3.12}
S_{E}(\rho\leq\rho_{s})&=2\int d\theta d\varphi\int_{0}^{\rho_{s}}d\tau_{E}\sqrt{g}\big\{\frac{-1}{\rho^{2}}+V(\phi)\big\}\\&=\int d\theta d\varphi\int_{0}^{\rho_{s}}d\tau_{E} \sqrt{g}\mathscr{L}_{eff}(\phi).
\end{split}
\end{eqnarray}
Here,
\begin{equation}
\label{eq:3.12m}
\mathscr{L}_{eff}(\phi)\equiv 2\big\{\frac{-1}{\rho^{2}}+V(\phi)\big\}
\end{equation}
is defined as the effective Lagrangian density.  Since the action $S_{E}(\rho\leq\rho_{s})$  in equation \eqref{eq:3.4} is valid in the spherically symmetric region $\rho\leq\rho_{s}$, the dynamical equations \eqref{eq:3.5}, \eqref{eq:3.6} and the Euclidean action \eqref{eq:3.12} are also valid in this region. More generally, regardless of the genus of the spacetime, if a given region is spherically symmetric, equations \eqref{eq:3.5}, \eqref{eq:3.6}  and \eqref{eq:3.12} hold in that region. In equation \eqref{eq:3.12}, the Euclidean action does not depend on the terms $\dot{\rho}$ and $\dot{\phi}$. Thus, equation \eqref{eq:3.12} is particularly convenient for calculating the total tunneling action $B$.

Denoting the effective Lagrangian density of the system in the region $\rho>\rho_{s}$ as $\mathscr{J}_{eff}(\phi)$. We assume that the coordinates ($x_{1}'$, $x_{2}'$, $x_{3}'$) are well defined in this region. Then, the Euclidean action of the system for the region $\rho>\rho_{s}$ can be formally expressed as
\begin{equation}
\label{eq:3.12m1}
S_{E}(\rho>\rho_{s})=\int_{\rho>\rho_{s}} dx_{1}'dx_{2}'dx_{3}'\sqrt{g'}\mathscr{J}_{eff}(\phi).
\end{equation}
Here, we use $S_{E}(\rho>\rho_{s})$ to denote the Euclidean action for the region $\rho>\rho_{s}$. The symbol $\int_{\rho>\rho_{s}}$ represents that the integration is performed over the region $\rho>\rho_{s}$. The factor $\sqrt{g'}$ represents the square root of the determinant of the spacetime metric in the coordinates ($x_{1}'$, $x_{2}'$, $x_{3}'$). The specific definition of the coordinates ($x_{1}'$, $x_{2}'$, $x_{3}'$) is not easy in high genus spacetimes~\cite{KB}. Fortunately, the exact form of  ($x_{1}'$, $x_{2}'$, $x_{3}'$) is not important for this study.

For convenience, we define
\begin{equation}
\label{eq:3.11}
B_{seed}\equiv B_{in} +B_{wall}
\end{equation}
as the vacuum decay seed.  If there are no wormhole fluctuations, $B=B_{seed}$. This is the characteristic of the conventional Coleman-De Luccia theory~\cite{SF}. However, if wormhole fluctuations are present, $B_{out}\neq 0$, and thus $B\neq B_{seed}$. To illustrate this result, we initially categorize wormhole fluctuations into three classes. In the first class, both ends of the wormhole are inside the bubble. In the second class, one end of the wormhole is inside the bubble, while the other end is outside the bubble. In the third class, both ends of the wormhole are outside the bubble. Intuitively, boundary wormholes belong to the third class, as depicted in figures~\ref{fig:m3} and~\ref{fig:m4}.

The size of the bubble is negligible compared to the size of the AdS$_{3}$ spacetime. Thus, the predominant wormhole fluctuations belong to the third class. Consequently, the first and second classes can be neglected.
Obviously, the third class of wormhole fluctuations  affects the  volume outside the bubble, thereby impacting the tunneling action $B_{out}$. In order to calculate $B_{out}$, recalling that in the case where there are no wormhole fluctuations, the tunneling action $B_{out}$ is defined as~\cite{SF}
\begin{eqnarray}\begin{split}
\label{eq:3.13}
B_{out}=\int_{\rho>\bar{\rho}; \ g= 0} d\theta d\varphi d\tau_{E}\sqrt{g}\mathscr{L}_{eff}(\phi_{bounce})-\int_{\rho>\bar{\rho};\  g= 0} d\theta d\varphi d\tau_{E}\sqrt{g}\mathscr{L}_{eff}(\phi_{2}).
\end{split}
\end{eqnarray}
Here, the symbol $\int_{\rho>\bar{\rho}; \ g= 0}$ represents that the integration is performed over the exterior of the bubble. $g=0$ represents that the genus of the spacetime is zero. The parent universe is spherically symmetric, ensuring that the integral in equation \eqref{eq:3.13} is well defined. On the right hand side of equation \eqref{eq:3.13}, the first term and the second term correspond to the bounce solution and the background term, respectively. Outside the bubble, $\phi_{bounce}=\phi_{2}$.  Thus, in the absence of wormhole fluctuations, $B_{out}=0$.

Taking BWFs into consideration, it is convenient to divide the region outside the bubble into two parts: $\overline{\rho}\leq\rho\leq\rho_{s}$ and $\rho>\rho_{s}$, as shown in figure~\ref{fig:m4}. As a result, the tunneling action $B_{out}$ can be written as
\begin{equation}
\label{eq:3.14}
B_{out}=B_{out}(\overline{\rho}\leq\rho\leq\rho_{s})+B_{out}(\rho>\rho_{s}).
\end{equation}
Here,
\begin{equation}
\label{eq:3.14m1}
B_{out}(\overline{\rho}\leq\rho\leq\rho_{s})\equiv\int d\theta d\varphi\int_{\overline{\rho}}^{\rho_{s}}d\tau_{E}\sqrt{g} \mathscr{L}_{eff}(\phi_{bounce}=\phi_{2})-\int d\theta d\varphi\int_{\overline{\rho}}^{\rho_{s}}d\tau_{E}\sqrt{g}\mathscr{L}_{eff}(\phi_{2}).
\end{equation}
On the right hand side of equation \eqref{eq:3.14},  $B_{out}(\overline{\rho}\leq\rho\leq\rho_{s})$ and $B_{out}(\rho>\rho_{s})$ represent the tunneling actions in the regions $\overline{\rho}\leq\rho\leq\rho_{s}$ and $\rho>\rho_{s}$, respectively. On the right hand side of equation \eqref{eq:3.14m1}, the first term corresponds to the bounce solution, while the second term is the background term. Note that in the region $\overline{\rho}\leq\rho\leq\rho_{s}$, there are no wormhole fluctuations. Thus, the tunneling action $B_{out}(\overline{\rho}\leq\rho\leq\rho_{s})=0$. This indicates that $B_{out}=B_{out}(\rho>\rho_{s})$.

Under the thin wall approximation, the bubble radius  $\bar{\rho}\gg0$~\cite{SF}. Thus, when $\rho>\bar{\rho}$, it can be expected  that $1/\rho^{2}\rightarrow 0$. This indicates that in the region $\overline{\rho}\leq\rho\leq\rho_{s}$, the effective Lagrangian density of the system  can be approximated as
\begin{equation}
\label{eq:3.14m2}
\mathscr{L}_{eff}(\phi)= 2V(\phi_{2}).
\end{equation}
Equation \eqref{eq:3.14m2} shows that in the region $\overline{\rho}\leq\rho\leq\rho_{s}$, the effective Lagrangian density is constant and  independent of the spacetime coordinates. In addition, outside the bubble, the scalar field is in  the state $V(\phi_{2})$. Thus, every point outside the bubble has the same cosmological constant. Therefore, at any point outside the bubble (such as points $a$, $b$ and $c$ in figure~\ref{fig:m4}), the effective Lagrangian density of the system should be described by equation \eqref{eq:3.14m2}. This implies that $\mathscr{L}_{eff}(\phi)=\mathscr{J}_{eff}(\phi)$. Consequently, the Euclidean action $S_{E}(\rho>\rho_{s})$ in equation \eqref{eq:3.12m1} becomes
\begin{equation}
\label{eq:3.14m2m}
S_{E}(\rho>\rho_{s})=2V(\phi_{2})\int_{\rho>\rho_{s}} dx_{1}'dx_{2}'dx_{3}'\sqrt{g'}.
\end{equation}
It is clear that the factor $\int_{\rho>\rho_{s}} dx_{1}'dx_{2}'dx_{3}'\sqrt{g'}$ in equation \eqref{eq:3.14m2m} represents the volume of the region $\rho>\rho_{s}$.

For convenience, we use $g=n$ to denote that the spacetime genus is equal to $n$  ($n=0, 1, 2, 3...$ ). For example, $g=0$ represents that the spacetime genus is zero.   We use $Vol(g=n;\rho>\rho_{s})$ to denote the volume of the region $\rho>\rho_{s}$ in a spacetime with genus $g=n$. Then, the Euclidean action $S_{E}(\rho>\rho_{s})$ in equation \eqref{eq:3.14m2m} can be rewritten as
\begin{equation}
\label{eq:3.14m3}
S_{E}(\rho>\rho_{s})=2V(\phi_{2})Vol(g=n;\rho>\rho_{s}).
\end{equation}
In equation \eqref{eq:3.14m3}, we have assumed that the spacetime genus produced by BWFs is $g=n$.

Combining equations \eqref{eq:3.8} and \eqref{eq:3.14m3}, the tunneling action $B_{out}(\rho>\rho_{s})$ can be written as
\begin{equation}
\label{eq:3.14m4}
B_{out}(\rho>\rho_{s})=2V(\phi_{2})Vol(g=n;\rho>\rho_{s})-2V(\phi_{2})Vol(g=0;\rho>\rho_{s}).
\end{equation}
Here, $Vol(g=0;\rho>\rho_{s})$ represents the volume of the region $\rho>\rho_{s}$ in the parent universe (the genus of the parent universe is zero).  On the right hand side of equation \eqref{eq:3.14m4}, the term $2V(\phi_{2})Vol(g=n;\rho>\rho_{s})$ represents the Euclidean action of the bounce solution in the region $\rho>\rho_{s}$, which depends on the volume of the genus $g=n$ spacetime. This term reflects the case where vacuum decay occurs in a genus $g=n$ spacetime. The term $2V(\phi_{2})Vol(g=0;\rho>\rho_{s})$ is the background term, corresponding to a solution where the system remains in its initial state. Thus,  the background term corresponds to the parent universe. Therefore, it is related to the volume of the parent universe.

We have illustrated that the region $\rho<\bar{\rho}$ is not altered by the third class of wormhole fluctuations (the first and second classes can be neglected). This indicates that
\begin{equation}
\label{eq:3.14m5}
Vol(g=n;\rho\leq\rho_{s})=Vol(g=0;\rho\leq\rho_{s}).
\end{equation}
Here, $Vol(g=n;\rho\leq\rho_{s})$ represents the volume of the region $\rho\leq\rho_{s}$ in a spacetime with genus $g=n$, while $Vol(g=0;\rho\leq\rho_{s})$ is the corresponding volume in the parent universe.
Combining equations \eqref{eq:3.14m4} and \eqref{eq:3.14m5}, the tunneling action $B_{out}$  can be written as
\begin{equation}
\label{eq:3.16}
B_{out}=B_{out}(\rho>\rho_{s})=2V(\phi_{2})\big\{Vol(g=n)-Vol(g=0) \big\}.
\end{equation}
In equation \eqref{eq:3.16}, $Vol(g=n)$ represents the volume of the AdS$_{3}$ spacetime with genus $g=n$, while $Vol(g=0)$ is the volume of the parent universe.

Both $Vol(g=n)$ and $Vol(g=0)$ are infinite. Typically, the regularized volume is used to replace these terms~\cite{KK}. We denote the regularized volume of the AdS$_{3}$ spacetime with genus $g=n$ as  $Rvol(g=n)$. Then, equation \eqref{eq:3.16} is modified to
\begin{equation}
\label{eq:3.17}
B_{out}=2V(\phi_{2})\big\{Rvol(g=n)-Rvol(g=0) \big\}.
\end{equation}
Equation \eqref{eq:3.17} implies that BWFs result in $B_{out}\neq 0$.

In this study, we only focus on physics at zero temperature.  Physically, as the temperature ($T$) of the thermal  AdS$_{3}$ spacetime (TAdS$_{3}$) approaches zero, the TAdS$_{3}$ spacetime transitions into a zero temperature AdS$_{3}$ spacetime with a conformal boundary shaped like a cylinder, as shown in figure~\ref{fig:m1}(a). Therefore, it is reasonable to consider that $\lim\limits_{T\rightarrow 0}$TAdS$_{3}$ is equivalent to the genus zero parent universe depicted in figure~\ref{fig:m1}(a). Hence, for convenience, we use $\lim\limits_{T\rightarrow 0}$TAdS$_{3}$ to define the genus zero parent universe.  Furthermore, we denote $V(\phi, T)$ as the scalar potential in the TAdS$_{3}$ (noting that temperature may influence the scalar potential) and set $\lim\limits_{T\rightarrow 0}V(\phi, T)$ to be equal to the scalar potential $V(\phi)$  in equation \eqref{eq:3.1}.

The genus of TAdS$_{3}$ is one~\cite{KK}.  We use $Rvol_{T}(g=1)$ to denote the regularized volume of TAdS$_{3}$. Then, the term $2V(\phi_{2})Rvol(g=0)$ in equation \eqref{eq:3.17} can be expressed as
\begin{equation}
\label{eq:3.18}
2V(\phi_{2})Rvol(g=0)=2\lim_{T\rightarrow 0}\Big(V(\phi_{2}, T)\cdot Rvol_{T}(g=1)\Big).
\end{equation}
Similarly, we use $Rvol_{T}(g=n+1)$ to represent the regularized volume of the spacetime where the parent universe is TAdS$_{3}$ and the genus generated by BWFs is $n$. And $g=n+1$ is the total genus of the AdS$_{3}$ handlebody.
Subsequently, the term $2V(\phi_{2})Rvol(g=n)$ in equation \eqref{eq:3.17} can be represented as
\begin{equation}
\label{eq:3.19}
2V(\phi_{2})Rvol(g=n)=2\lim_{T\rightarrow 0}\Big(V(\phi_{2}, T)\cdot Rvol_{T}(g=n+1)\Big).
\end{equation}
Hence,  equation \eqref{eq:3.17} can be rewritten as
\begin{equation}
\label{eq:3.20}
B_{out}=2\lim_{T\rightarrow 0}\Big\{V(\phi_{2}, T)\cdot\Big(Rvol_{T}(g=n+1)-Rvol_{T}(g=1) \Big)\Big\}.
\end{equation}
In equation \eqref{eq:3.20}, when $n=0$, $B_{out}=0$.  This corresponds to the case where there are no wormhole fluctuations.

The regularized volume of the AdS$_{3}$ handlebody can be obtained from the regularized Euclidean Einstein-Hilbert action of the AdS$_{3}$ spacetime,  defined as~\cite{HSB,KK}
\begin{equation}
\label{eq:3.21}
I_{reg}\equiv-\frac{1}{2}\int dx^{3}\sqrt{g}(R-2\Lambda)+2\int dx^{2}\sqrt{\gamma}(K-\sqrt{-\Lambda})+Count..
\end{equation}
Here, $\Lambda$ is the cosmological constant ($\Lambda<0$), while  $K$ and $\gamma$ represent the trace of the extrinsic curvature and  the induced metric on the boundary, respectively. The term $Count.$ refers to the counterterm used to remove the logarithmic divergence~\cite{KK}. A detailed discussion of this counterterm  can be found in~\cite{HSB}. The regularized action \eqref{eq:3.21} and the Euclidean Einstein-Hilbert action $-\int dx^{3}\sqrt{g}(R/2-\Lambda)$ correspond to the same dynamical equation.

In~\cite{KK}, Krasnov demonstrated that the regularized action $I_{reg}$ can be simplified as
\begin{equation}
\label{eq:3.22}
I_{reg}=2\pi\sqrt{\frac{-1}{\Lambda}}(\mathrm{ln}|k_{1}|+\mathrm{ln}\rho_{2}\rho'_{2}\cdot\cdot\cdot\rho_{g}\rho'_{g}),
\end{equation}
where $\rho_{i}$ ($\rho'_{i}$) represents the radius of the Schottky circle $C_{i}$ ($C'_{i}$), and $k_{1}$ is the multiplier of the generator $L_{1}$ of the Schottky group $\mathbf{Sch}(L_{1}, L_{2}, ..., L_{g} )$. Definitions of the Schottky circle, Schottky group, and the multiplier $k_{1}$ of the generator $L_{1}$ are provided in appendix~\ref{sec:A0}.
In the case of $g=1$, the regularized action is $I_{reg}=2\pi\sqrt{-1/\Lambda}\mathrm{ln}|k_{1}|$. This is the regularized action of TAdS$_{3}$~\cite{KK}. Thus, the variable $k_{1}$ and the generator $L_{1}$ are associated with TAdS$_{3}$. The action of the BTZ black hole can also be represented as the regularized action $I_{reg}$ with $g=1$~\cite{KK}. The distinction between TAdS$_{3}$ and the BTZ black hole lies in the fact that, for TAdS$_{3}$, the time circle (a circle where different points on the circle correspond to different times) is non contractible, whereas, for the BTZ black hole, the time circle is contractible~\cite{KK}.

Equation \eqref{eq:3.22} can also be derived from the Takhtajan-Zograf  action, which is defined as~\cite{HSB,KK,LTPZ}
\begin{eqnarray}\begin{split}
\label{eq:3.22a}
I_{TZ}=&\frac{1}{2}\sqrt{\frac{-1}{\Lambda}}\Big\{\iint_{F_{S}}\frac{i}{2}dz\wedge d\bar{z}(4\partial_{z}\psi\partial_{\bar{z}}\psi+\frac{1}{2}e^{2\psi})+i\int_{C_{j}}\sum_{j=2}^{g}(d\bar{z}\psi\frac{\partial^{2}_{z}\bar{L}_{j}}{\partial_{z}\bar{L}_{j}}-dz\psi\frac{\partial^{2}_{z}L_{j}}{\partial_{z}L_{j}}) \\& +\frac{i}{2} \int_{C_{j}}\sum_{j=2}^{g}\mathrm{ln}|\partial_{z}L_{j}|^{2}\frac{\partial^{2}_{z}L_{j}}{\partial_{z}L_{j}}dz+4\pi\sum_{j=2}^{g}\mathrm{ln}|(L_{j})_{21}|^{2} \Big\}.
\end{split}
\end{eqnarray}
Here, $I_{TZ}$ and $\psi$ represent the Takhtajan-Zograf  action and the Liouville conformal field, respectively.  $C_{j}$ and $F_{S}$  represent the Schottky circle and the fundamental domain of the Schottky group $\mathbf{Sch}(L_{1}, L_{2}, ..., L_{g} )$ acting on the Riemann sphere $\mathbb{S}$, respectively. $\bar{z}$ and $\bar{L}_{j}$ represent the complex conjugate of $z$ and $L_{j}$, respectively. $(L_{j})_{21}$ refers to the element in the second row and  first column of the generator $L_{j}$. Equation \eqref{eq:2.3b} in appendix~\ref{sec:A0} indicates that $(L_{j})_{21}=(1-k_{j})/(\sqrt{|k_{j}|}|\xi_{j}-\eta_{j}|)$. Here, $ \xi_{j}, \eta_{j}$ and $ k_{j}$ are the attractive fixed point, repelling fixed point and multiplier of the generator $L_{j}$, respectively. Detailed explanations  of these parameters are provided in appendix~\ref{sec:A0}. The Takhtajan-Zograf action  corresponds to the action of the Liouville conformal field theory defined on the conformal boundary of the AdS$_{3}$ handlebody spacetime. One can also use holographic duality to study wormhole fluctuations and AdS vacuum decay~\cite{XY1,AE,TG22,JE22,DL22,JM22}. However, this approach is beyond the scope of this work and will not be discussed further.

The relationship between the radius of the Schottky circle and the parameters ($ \xi_{i}, \eta_{i}, k_{i}$) is given by~\cite{AS,BM}
\begin{equation}
\label{eq:3.23}
\rho_{i}=\rho'_{i}=\big|\frac{\sqrt{k_{i}}(\xi_{i}-\eta_{i})}{1-k_{i}}\big|.
\end{equation}
Therefore, the regularized action $I_{reg}$ can be expressed as
\begin{equation}
\label{eq:3.24}
I_{reg}=2\pi\sqrt{\frac{-1}{\Lambda}}\Big\{\mathrm{ln}|k_{1}|+2\mathrm{ln}\big|\frac{\sqrt{k_{2}}(\xi_{2}-\eta_{2})}{1-k_{2}}\frac{\sqrt{k_{3}}(\xi_{3}-\eta_{3})}{1-k_{3}}\cdot\cdot\cdot\frac{\sqrt{k_{g}}(\xi_{g}-\eta_{g})}{1-k_{g}}\big|\Big\}.
\end{equation}
The normalized condition of the Schottky group implies that $\xi_{2}=1$. Equation \eqref{eq:3.24} clearly indicates that the regularized action $I_{reg}$ is a $(6g-6)$-variate function in the Schottky space. The variables are $(k_{1}, \eta_{2}, k_{2}, \xi_{3}, \eta_{3}, k_{3},\cdot\cdot\cdot, \xi_{g}, \eta_{g}, k_{g})$ with $g\geq 2$. In the case of $g=1$, the unique variable of $I_{reg}$ is $k_{1}$. The variable $k_{1}$ describes the properties of TAdS$_{3}$, while the other variables $(\eta_{2}, k_{2}, \xi_{3}, \eta_{3}, k_{3},\cdot\cdot\cdot, \xi_{g}, \eta_{g}, k_{g})$ describe the properties of the boundary wormholes.

Recalling that for the TAdS$_{3}$, the relationship between the regularized action   and the regularized volume $Rvol_{T}(g=1)$ is given by $I_{reg}=-2\Lambda \cdot Rvol_{T}(g=1)$. This relationship can be easily extended to any genus AdS$_{3}$ spacetime: $I_{reg}=-2\Lambda \cdot Rvol(g=n)$. Therefore, based on equation \eqref{eq:3.24}, the regularized volume $Rvol_{T}(g=n+1)$ can be represented as
\begin{eqnarray}\begin{split}
\label{eq:3.25}
Rvol_{T}(g=n+1)=-&\frac{\pi}{\Lambda}\sqrt{\frac{-1}{\Lambda}}\Big\{\mathrm{ln}|k_{1}|+2\mathrm{ln}\big|\frac{\sqrt{k_{2}}(1-\eta_{2})}{1-k_{2}}\big|\\&\ +2\mathrm{ln}\big|\frac{\sqrt{k_{3}}(\xi_{3}-\eta_{3})}{1-k_{3}}\cdot\cdot\cdot\frac{\sqrt{k_{g}}(\xi_{g}-\eta_{g})}{1-k_{g}}\big|\Big\}.
\end{split}
\end{eqnarray}
The term $-\pi\Lambda^{-1}\sqrt{-\Lambda^{-1}}\mathrm{ln}|k_{1}|$ in equation \eqref{eq:3.25} represents the regularized volume of TAdS$_{3}$.

Assuming that the period matrix of  TAdS$_{3}$ is $\Pi(1)$. For the genus one handlebody, $\Pi(1)$ corresponds to the modular parameter of the torus (refer to appendix~\ref{sec:A1}). Equation \eqref{eq:2.16} in appendix~\ref{sec:A1} shows that $2\pi i\Pi(1)=\mathrm{ln}k_{1}$. Furthermore, the inverse of the temperature ($\frac{1}{T}$) of TAdS$_{3}$ is proportional to the imaginary part of $\Pi(1)$~\cite{KK,SAX}. Therefore, $\frac{1}{T}\propto \mathrm{ln}|k_{1}|$. Other parameters $( \eta_{2}, k_{2}, \xi_{3}, \eta_{3}, k_{3},\cdot\cdot\cdot, \xi_{g}, \eta_{g}, k_{g})$ are independent of the multiplier $k_{1}$, and thus they are also independent of the temperature $T$.  The term proportional to $\mathrm{ln}|k_{1}|$ in equation \eqref{eq:3.25} is cancelled by the term $Rvol_{T}(g=1)$ in equation \eqref{eq:3.20}.
Combining equation \eqref{eq:3.25} with this conclusion, equation \eqref{eq:3.20} can be written as
\begin{equation}
\label{eq:3.26}
B_{out}=-4\pi \sqrt{\frac{-1}{V(\phi_{2})}}\Big\{\mathrm{ln}\big|\frac{\sqrt{k_{2}}(1-\eta_{2})}{1-k_{2}} \frac{\sqrt{k_{3}}(\xi_{3}-\eta_{3})}{1-k_{3}}\cdot\cdot\cdot\frac{\sqrt{k_{g}}(\xi_{g}-\eta_{g})}{1-k_{g}}\big|\Big\}.
\end{equation}
Here, we have used the fact that in the parent universe $\lim\limits_{T\rightarrow 0}$TAdS$_{3}$, the cosmological constant is equal to $V(\phi_{2})$ (outside the bubble). Equation \eqref{eq:3.26} indicates that the tunneling action $B_{out}$ is a $(6g-8)$-variate function in the Schottky space.

In addition, the third class of wormhole fluctuations is independent of the bounce bubble, meaning it does not affect the vacuum decay seed $B_{seed}$. Therefore, when calculating the vacuum decay seed $B_{seed}$, one can simply assume that there are no wormhole fluctuations. Using the method described in reference~\cite{SF}, one can obtain
\begin{eqnarray}\begin{split}
\label{eq:3.27}
B_{seed}=4\pi\Big\{&V^{-\frac{1}{2}}(\phi_{2})\ \mathrm{arcsin}(\bar{\rho}\sqrt{V(\phi_{2})})-V^{-\frac{1}{2}}(\phi_{1})\ \mathrm{arcsin}(\bar{\rho}\sqrt{V(\phi_{1})})\\&+\bar{\rho}\sqrt{1-\bar{\rho}^{2}V(\phi_{2})}-\bar{\rho}\sqrt{1-\bar{\rho}^{2}V(\phi_{1})}+\bar{\rho}^{2}\mu\Big\}
\end{split}
\end{eqnarray}
with the radius
\begin{equation}
\label{eq:3.28}
\bar{\rho}=\frac{2\mu}{\sqrt{\mu^{4}+\epsilon^{2}+2\mu^{2}(2V(\phi_{2})+\epsilon)}}.
\end{equation}
Here, $\mu\equiv\int_{\phi_{1}}^{\phi_{2}}d\phi\sqrt{2(V(\phi)-V(\phi_{2}))}$ represents the tension of the domain wall~\cite{HJ,VD}. The detailed derivation of equations \eqref{eq:3.27} and \eqref{eq:3.28} is presented in appendix~\ref{sec:A}.

We point out that the parameters $( \eta_{2}, k_{2}, \xi_{3}, \eta_{3}, k_{3},\cdot\cdot\cdot, \xi_{g}, \eta_{g}, k_{g})$  are defined in Schottky space (see appendix~\ref{sec:A0}), while the coordinates ($x_{1}'$, $x_{2}'$, $x_{3}'$) are defined in Euclidean AdS$_{3}$ spacetime. Although equation \eqref{eq:3.12m1} involves the coordinates ($x_{1}'$, $x_{2}'$, $x_{3}'$), equations \eqref{eq:3.26} and \eqref{eq:3.27} clearly show that both the tunneling action $B_{out}$ and the vacuum decay seed $B_{seed}$ are independent of these coordinates. Thus, the total tunneling action $B$ does not depend on the coordinates ($x_{1}'$, $x_{2}'$, $x_{3}'$). The calculations in the later sections are based on equations \eqref{eq:3.26} and \eqref{eq:3.27}. Therefore,  the exact form of the coordinates ($x_{1}'$, $x_{2}'$, $x_{3}'$) is not essential for this study.

To sum up, in this section, we generalize the tunneling action to the case where there are BWFs. The main contribution comes from the third class of wormhole fluctuations, which does not affect the vacuum decay seed $B_{seed}$.
The relationship between the total tunneling action $B$ and the vacuum decay seed $B_{seed}$ is given by $B=B_{seed}+B_{out}$. In the absence of wormhole fluctuations, $B_{out}=0$. When BWFs are present, $B_{out}$ is described by equation \eqref{eq:3.26}.

\subsection{Representing the  tunneling action in the Siegel upper half space}
\label{sec:3.2}

The parameters $(\eta_{2}, k_{2}, \xi_{3}, \eta_{3}, k_{3},\cdot\cdot\cdot, \xi_{g}, \eta_{g}, k_{g})$ in equation \eqref{eq:3.26} specify the BWFs. To include the contributions from all BWFs (the third class), one needs to integrate over these parameters. However, the Schottky space is not simply connected~\cite{KK},  and different points in the Schottky space may correspond to the same BWFs. This makes it inconvenient to define the integral domain directly in the Schottky space. Instead, we perform the summation of the BWFs in the Siegel upper half space ($\mathbf{Sie}(g)$). By using the relationship between the elements of the period matrix and the parameters $(\xi_{i}, \eta_{i}, k_{i})$, the tunneling action can be transformed from the Schottky space to $\mathbf{Sie}(g)$. Detailed information about the Siegel upper half space and the period matrix is provided in appendix~\ref{sec:A1}.

Combining the normalized condition ($\xi_{1}=0$, $\xi_{2}=1$, $\eta_{1}=\infty$) of the Schottky group  and equation \eqref{eq:2.17} in appendix~\ref{sec:A1}, one can obtain
\begin{equation}
\label{eq:3.29}
\eta_{2}=\mathrm{exp}(-2\pi i \Pi(g)_{12}),
\end{equation}
\begin{equation}
\label{eq:3.30}
\xi_{j}=\eta_{j}\mathrm{exp}(2\pi i \Pi(g)_{1j}),
\end{equation}
and
\begin{equation}
\label{eq:3.31}
2\pi i \Pi(g)_{2j}=\mathrm{ln}\frac{(\eta_{2}-\eta_{j})(1-\xi_{j})}{(\eta_{2}-\xi_{j})(1-\eta_{j})}.
\end{equation}
In equations \eqref{eq:3.29}-\eqref{eq:3.32}, $\Pi(g)_{ij}$ denotes the element of the period matrix (refer to appendix~\ref{sec:A1}).

Substituting equations \eqref{eq:3.29} and \eqref{eq:3.30} into equation \eqref{eq:3.31}, one can obtain the relationship between the repelling fixed point $\eta_{j}$ and the elements of the period matrix as
\begin{equation}
\label{eq:3.32}
\eta_{j}=\frac{-\Delta_{2j}\pm \sqrt{\Delta_{2j}^{2}-4\Delta_{1j}\Delta_{3j}}}{2\Delta_{1j}}.
\end{equation}
Here, the symbol $``\pm"$ corresponds to two different results. In the later, we will show that only one of the results is meaningful in our model (see appendix~\ref{sec:B}).  The terms $\Delta_{1j}$, $\Delta_{2j}$, and $\Delta_{3j}$ are defined as follows:
\begin{equation}
\label{eq:3.33}
\Delta_{1j}\equiv \mathrm{exp}(2\pi i \Pi(g)_{2j})-\mathrm{exp}(2\pi i \Pi(g)_{1j}),
\end{equation}
\begin{eqnarray}\begin{split}
\label{eq:3.34}
\Delta_{2j}\equiv &1-\mathrm{exp}(2\pi i \Pi(g)_{2j})\mathrm{exp}(-2\pi i \Pi(g)_{12})-\mathrm{exp}(2\pi i \Pi(g)_{1j})\mathrm{exp}(2\pi i \Pi(g)_{2j})\\&+\mathrm{exp}(-2\pi i \Pi(g)_{12})\mathrm{exp}(2\pi i \Pi(g)_{1j}),
\end{split}
\end{eqnarray}
\begin{equation}
\label{eq:3.35}
\Delta_{3j}\equiv \mathrm{exp}(2\pi i \Pi(g)_{2j})\mathrm{exp}(-2\pi i \Pi(g)_{12})-\mathrm{exp}(-2\pi i \Pi(g)_{12}).
\end{equation}
Equation \eqref{eq:3.32} is the representation of the repelling fixed point $\eta_{j}$ in the space $\mathbf{Sie}(g)$.

Equation \eqref{eq:2.16} in appendix~\ref{sec:A1} indicates that $k_{i}=\mathrm{exp}(2 \pi i \Pi(g)_{ii})$. Using this result and  incorporating equations \eqref{eq:3.30} and \eqref{eq:3.32} into equation \eqref{eq:3.26}, the tunneling action $B_{out}$ becomes
\begin{eqnarray}\begin{split}
\label{eq:3.36}
B_{out}=\frac{-4\pi}{\sqrt{-V(\phi_{2})}}\Big\{&\sum_{j=2}^{g}\mathrm{ln}\Big|\frac{\mathrm{exp}(\pi i \Pi(g)_{jj})(\mathrm{exp}(2 \pi i \Pi(g)_{1j})-1)}{1-\mathrm{exp}(2\pi i \Pi(g)_{jj})}\Big|\\&+\sum_{j=2}^{g}\mathrm{ln}\Big|\frac{-\Delta_{2j}\pm \sqrt{\Delta_{2j}^{2}-4\Delta_{1j}\Delta_{3j}}}{2\Delta_{1j}}\Big|\Big\}.
\end{split}
\end{eqnarray}
Equations \eqref{eq:3.32}-\eqref{eq:3.36} show that in the space $\mathbf{Sie}(g)$, the tunneling action $B_{out}$ depends on the set of parameters $(\Pi(g)_{22}, \Pi(g)_{33}, \cdot\cdot\cdot, \Pi(g)_{gg}; \Pi(g)_{12}, \Pi(g)_{13}, \cdot\cdot\cdot, \Pi(g)_{1g}; \Pi(g)_{23}, \Pi(g)_{24}, \cdot\cdot\cdot, \Pi(g)_{2g})$. These parameters fully specify the BWFs.

In our setup,  the generator $L_{1}$ of the Schottky group $\mathbf{Sch}(L_{1}, L_{2}, ..., L_{g} )$,  the multiplier $k_{1}$, and the parameter $\Pi(g)_{11}$ ($2\pi i\Pi(g)_{11}=\mathrm{ln}k_{1}$) all correspond to the parent universe $\lim\limits_{T\rightarrow 0}$TAdS$_{3}$. Other generators $L_{i}$ ($i\neq 1$) correspond to the wormholes. Thus, in equation \eqref{eq:3.36}, $j\geq 2$. The maximum value of $j$ being $g$ implies that the genus produced by the BWFs is $g-1$. The element $\Pi(g)_{1j}$ ($j\neq 1$) should be interpreted as the interaction between the parent universe $\lim\limits_{T\rightarrow 0}$TAdS$_{3}$ and wormhole $j$. And the element $\Pi(g)_{ij}$ ($i, j\neq 1$; $i\neq j$) should be interpreted as the interaction between  wormholes $i$ and $j$ (refer to appendix~\ref{sec:A1} or reference~\cite{AS}).

To simplify equation \eqref{eq:3.36}, we denote $\Pi(g)=X+iY$, where $X$ and $Y$ are real symmetric $g\times g$ matrices. Define
\begin{equation}
\label{eq:3.43}
\Xi(\phi_{2})\equiv 2\pi \sqrt{\frac{-1}{V(\phi_{2})}}.
\end{equation}
Under the dilute wormhole gas approximation~\cite{SK,AS} (refer to appendix~\ref{sec:A1}),  equation \eqref{eq:3.36} can be simplified to
\begin{eqnarray}\begin{split}
\label{eq:3.44}
B_{out}=&-\Xi(\phi_{2})(g-1)\mathrm{ln}\Big(1+\mathrm{exp}(4\pi Y_{12})-2\mathrm{cos}(2\pi X_{12})\mathrm{exp}(2\pi Y_{12})\Big)\\&-\Xi(\phi_{2})\sum_{j=2}^{g}\Big\{-2\pi Y_{jj}-\mathrm{ln}\Big(1+\mathrm{exp}(-4\pi Y_{jj})-2\mathrm{cos}(2\pi X_{jj})\mathrm{exp}(-2\pi Y_{jj})\Big)\\&\ \ \ \ \ \ \ \ \    +\mathrm{ln}\Big(1+\mathrm{exp}(-4\pi Y_{1j})-2\mathrm{cos}(2\pi X_{1j})\mathrm{exp}(-2\pi Y_{1j})\Big)\Big\}.
\end{split}
\end{eqnarray}
The derivation of equation \eqref{eq:3.44} is presented in appendix~\ref{sec:B}. Equation \eqref{eq:3.44} shows that the tunneling action $B_{out}$ is a function of the set of variables $(X_{22}, X_{33}, \cdot\cdot\cdot, X_{gg}, Y_{22}, Y_{33}, \cdot\cdot\cdot, Y_{gg}, X_{12}, X_{13}, \cdot\cdot\cdot, X_{1g}, Y_{12}, Y_{13}, \cdot\cdot\cdot, Y_{1g})$.

The tunneling action $B_{out}$ in equation \eqref{eq:3.44} is still complicated.  In appendix~\ref{sec:B}, we have demonstrated that under the dilute wormhole gas approximation, the interactions  between different wormholes are very weak, giving justification to ignore them. To further simplify equation \eqref{eq:3.44} , we consider the case where the interactions between the parent universe and the wormholes are also small. Specifically, we focus on the case where
\begin{equation}
\label{eq:3.45}
\Pi(g)_{ij} \ \ (i, j\neq 1; i\neq j)\ll \Pi(g)_{1j}\ \  (j\neq 1)\ll 1.
\end{equation}
This implies that all non-diagonal elements of the period matrix are small,  and the interactions between different wormholes  are significantly weaker than those between the parent universe and the wormholes.

Under condition \eqref{eq:3.45}, it is reasonable to neglect the interactions between different wormholes and approximate the terms related to the interactions between the parent universe and the wormholes to the first order. Thus, $\mathrm{cos}(2\pi X_{1j})\approx 1$ and $\mathrm{exp}(\pm 2\pi Y_{1j})\approx 1\pm 2\pi Y_{1j}$. Using these approximations, equation \eqref{eq:3.44} can be simplified to
\begin{eqnarray}\begin{split}
\label{eq:3.46}
B_{out}=2\Xi(\phi_{2})\sum_{j=2}^{g}\big\{\pi Y_{jj}-\mathrm{ln}(2\pi Y_{1j})\big\} -2\Xi(\phi_{2})(g-1)\mathrm{ln}(2\pi Y_{12}).
\end{split}
\end{eqnarray}
The element $\Pi_{1j} (j\neq 1)$ represents the interaction between the parent universe and wormhole ``$j$". Equation \eqref{eq:3.46} indicates that  the tunneling action $B_{out}$ is independent of the variable $X_{1j}$. Thus, under condition \eqref{eq:3.45}, $Y_{1j} (j\neq 1)$ can be used to describe the coupling strength between the parent universe and wormhole ``$j$". Equation \eqref{eq:3.46}  expresses the tunneling action $B_{out}$ in the space  $\mathbf{Sie}(g)$.  Note that the vacuum decay seed $B_{seed}$ is independent of the elements of the period matrix. Therefore, by representing $B_{out}$ in the space $\mathbf{Sie}(g)$, the total tunneling action $B$ is also represented in the space $\mathbf{Sie}(g)$.

\section{Integration for the non-equivalent boundary wormhole fluctuations }
\label{sec:4}
In the Coleman-De Luccia theory, the vacuum decay rate $e^{-B_{seed}}$ corresponds to scenarios without wormhole fluctuations. However, during the vacuum decay process, spacetime can fluctuate to generate one or more wormholes. Various fluctuations correspond to different vacuum decay processes. To calculate the total vacuum decay rate, all possible vacuum decay processes must be taken into account. Different BWFs are represented by distinct handlebodies. Thus,  one needs to integrate over the non-equivalent handlebodies. The vacuum decay rate associated with a specific type of boundary wormhole fluctuation is $e^{-(B_{seed}+B_{out})}$, where $B_{out}$ is given by equation \eqref{eq:3.46}. Therefore, the total vacuum decay rate defined by equation \eqref{eq:3.2m} should be
\begin{equation}
\label{eq:4.1}
\Gamma_{tot}=e^{-B_{seed}}+\sum_{g=2}^{\infty}\int_{\mathfrak{F}(g)}e^{-(B_{seed}+B_{out})}d\sigma(g),
\end{equation}
where $\mathfrak{F}(g)$ and $d\sigma(g)$ represent the integral domain and the integral measure, respectively.

In equation \eqref{eq:4.1}, the first term on the right hand side represents the vacuum decay rate for the scenario without wormhole fluctuations. The second term accounts for the contributions from BWFs. The parameter $g$ starts at $g=2$ because $g=2$ represents the genus produced by the BWFs as one.  The vacuum decay seed $B_{seed}$ is given by equation \eqref{eq:3.27}. The total tunneling action $B_{seed}+B_{out}$ is determined by equations \eqref{eq:3.27} and \eqref{eq:3.46}.

In the integral domain $\mathfrak{F}(g)$, each point represents a different handlebody spacetime. In other words, each point in $\mathfrak{F}(g)$ corresponds to a different boundary wormhole fluctuation. To elucidate the integral domain $\mathfrak{F}(g)$, it is crucial to note  that  in the space $\mathbf{Sie}(g)$, when two period matrices are associated by a modular transformation, they describe conformally equivalent Riemann surfaces $\mathbf{\Sigma}_{g}$ ( $\mathbf{\Sigma}_{g}$ denotes the closed Riemann surface of genus $g$, see appendix~\ref{sec:A0} for more details ). The modular transformation between two period matrices $\Pi(g)^{(a)}$ and $\Pi(g)^{(b)}$ is defined as~\cite{M}
\begin{equation}
\label{eq:4.2}
\Pi(g)^{(a)}=\frac{A\Pi(g)^{(b)}+B}{C\Pi(g)^{(b)}+D},
\end{equation}
where $A$, $B$, $C$, $D$ are $g\times g$ matrices.  Each element of these matrices is an integer. These matrices must satisfy the conditions $AB^{T}=BA^{T}$, $CD^{T}=DC^{T}$, and $AD^{T}-BC^{T}=\mathbb{I}_{g}$~\cite{AE2,SA}, where $\mathbb{I}_{g}$ represents the $g\times g$ identity matrix and $M^{T}$ denotes the transpose of the matrix $M$.

The set of all modular transformations forms the modular group $\mathrm{Sp}(2g, \mathbb{Z})$. This group has two generators~\cite{HM}
\begin{equation}
\label{eq:4.3}
G(g)^{(1)}=
\begin{pmatrix}
\mathbb{I}_{g} & S\\
0 & \mathbb{I}_{g}\\
\end{pmatrix},
\ \ \ \ \ \ \ \ \
G(g)^{(2)}=
\begin{pmatrix}
0 & \mathbb{I}_{g}\\
-\mathbb{I}_{g} & 0\\
\end{pmatrix}.
\end{equation}
Here,  $S$ is an arbitrary $g\times g$ symmetric integer matrix. Every element of the modular group $\mathrm{Sp}(2g, \mathbb{Z})$
can be generated by these two generators.  Equations \eqref{eq:4.2} and \eqref{eq:4.3} show that the generator $G(g)^{(1)}$ acts on the period matrix $\Pi(g)$ as $\Pi(g)\rightarrow \Pi(g)+S$, while the generator $G(g)^{(2)}$ acts on $\Pi(g)$ as $\Pi(g)\rightarrow -\Pi(g)^{-1}$. Some researchers may claim that the modular group $\mathrm{Sp}(2g, \mathbb{Z})$ has three generators~\cite{HM,NL}. However, an additional generator can be generated by the generators $G(g)^{(1)}$ and $G(g)^{(2)}$~\cite{HM}. Therefore, it is sufficient to consider the generators $G(g)^{(1)}$ and $G(g)^{(2)}$.

\begin{figure}[tbp]
\centering
\includegraphics[width=12cm]{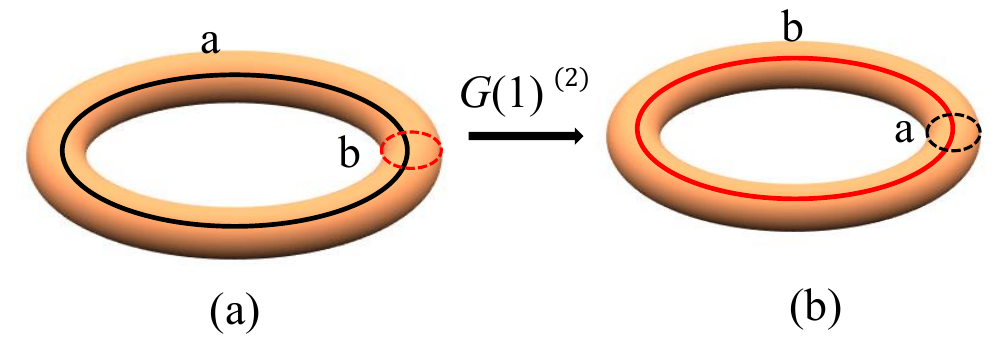}
\caption{\label{fig:5} The transformation $G(1)^{(2)}$ transforms the torus (a) into the torus (b). In figures (a) and (b), the tori have two dimensions $a$ and $b$. The transformation $G(1)^{(2)}$ exchanges the two dimensions $a$ and $b$.}
\end{figure}

The simplest case is the genus one closed Riemann surface $\mathbf{\Sigma}_{1}$. In this case, the period matrix $\Pi(1)$ is the modular parameter of the torus. The related modular group is $\mathrm{Sp}(2, \mathbb{Z})$, which is generated by the following generators~\cite{XG,FJ,GSX}
\begin{equation}
\label{eq:4.4}
G(1)^{(1)}=
\begin{pmatrix}
1 & 1\\
0 & 1\\
\end{pmatrix},
\ \ \ \ \ \ \ \ \
G(1)^{(2)}=
\begin{pmatrix}
0 & 1\\
-1 & 0\\
\end{pmatrix}.
\end{equation}
It is straightforward to verify that $G(1)^{(1)}$ and $G(1)^{(2)}$ satisfy the multiplication rules $(G(1)^{(2)})^{2}=(G(1)^{(2)}G(1)^{(1)})^{3}=\mathbb{I}_{2}$~\cite{XG}. The generator $G(1)^{(1)}$ generates the Dehn twist of the torus ($\Pi(1)\rightarrow \Pi(1)+1$). The transformations generated by $G(1)^{(2)}$ exchange the two dimensions of the torus ($\Pi(1)\rightarrow -1/\Pi(1)$)~\cite{MN}, as shown in figure~\ref{fig:5}.
If the torus $\Pi(1)$ represents a two dimensional spacetime, then the torus $-1/\Pi(1)$  also represents a two dimensional spacetime. Mathematically, the tori $\Pi(1)$ and $-1/\Pi(1)$ share the same conformal structure. However, physically, they represent different spacetimes due to differences in their time circles. Filling their interiors results in two non-equivalent three-dimensional handlebody spacetimes. Thus, the transformation $G(1)^{(2)}$ may alter the properties of the handlebody  spacetime. A specific example is that the modular transformation $G(1)^{(2)}$ converts the TAdS$_{3}$ spacetime into the BTZ black hole~\cite{AE}.

The initial (or final) state  defined in the TAdS$_{3}$ spacetime and the BTZ black hole corresponds to a spatial slice. The spatial slice of  TAdS$_{3}$  is topologically distinct from that of  the BTZ black hole. This indicates that one cannot define the same initial (or final) state in the TAdS$_{3}$ spacetime and the BTZ black hole. At least, the topological structures of the initial (or final) state  defined in these two types of spacetimes are different.
More generally, if the modular transformation $G(1)^{(2)}$   exchanges the time and space dimensions of a handlebody spacetime, it becomes impossible to define the same initial (or final) state  in the spacetimes associated by the transformation $G(1)^{(2)}$. Therefore, specifying an initial state implies selecting a specific spacetime from the set of spacetimes associated with the modular transformation $G(1)^{(2)}$.

The generators $G(g)^{(1)}$ and $G(g)^{(2)}$ are generalizations of the generators $G(1)^{(1)}$ and $G(1)^{(2)}$, respectively. Similar to the case of $G(1)^{(2)}$, one can easily show that $(G(g)^{(2)})^{2}=\mathbb{I}_{2g}$.   This implies that the modular transformations generated by $G(g)^{(2)}$, when acting on a genus $g$ handlebody spacetime, can only result in one other distinct spacetime.  The generator $G(g)^{(2)}$ acts on the period matrix as $\Pi(g)\rightarrow -\Pi(g)^{-1}$.

Under condition \eqref{eq:3.45}, the period matrix $\Pi(g)$ can be written as $\Pi(g)=\bigoplus\limits_{i=1}^{g}\Pi(g)_{ii}+o(\Pi(g)_{ij})$, where $o(\Pi(g)_{ij})$ is a small term determined by the off-diagonal elements of $\Pi$.  Utilizing the formula~\cite{FG}
\begin{equation}
\label{eq:4.5}
(A-B)^{-1}=A^{-1}+A^{-1}BA^{-1}+A^{-1}BA^{-1}BA^{-1}+\cdot\cdot\cdot
\end{equation}
the inverse of the period matrix $-\Pi(g)^{-1}$  can be written as
\begin{eqnarray}\begin{split}
\label{eq:4.6}
-\Pi(g)^{-1}=&-(\bigoplus_{i=1}^{g}\Pi(g)_{ii}^{-1})\Big\{\mathbb{I}_{g}-o(\Pi(g)_{ij})(\bigoplus_{i=1}^{g}\Pi(g)_{ii}^{-1})\\&\ \ \ \ \ +o(\Pi(g)_{ij})(\bigoplus_{i=1}^{g}\Pi(g)_{ii}^{-1})o(\Pi(g)_{ij})(\bigoplus_{i=1}^{g}\Pi(g)_{ii}^{-1})+\cdot\cdot\cdot\Big\}.
\end{split}
\end{eqnarray}
The dominant term on the right hand side of equation \eqref{eq:4.6} is $-\bigoplus\limits_{i=1}^{g}\Pi(g)_{ii}^{-1}$.
Hence, the leading terms of $\Pi(g)$ and $-\Pi(g)^{-1}$ are $\bigoplus\limits_{i=1}^{g}\Pi(g)_{ii}$ and $-\bigoplus\limits_{i=1}^{g}\Pi(g)_{ii}^{-1}$, respectively. Both $\bigoplus\limits_{i=1}^{g}\Pi(g)_{ii}$ and $-\bigoplus\limits_{i=1}^{g}\Pi(g)_{ii}^{-1}$ represent multiple disconnected spacetimes.

We have demonstrated that it is impossible to define the same initial (or final) state  in the spacetimes $\Pi(g)_{ii}$ and $-\Pi(g)_{ii}^{-1}$. Thus,  one also cannot define the same initial (or final) state in the spacetimes $\bigoplus\limits_{i=1}^{g}\Pi(g)_{ii}$ and $-\bigoplus\limits_{i=1}^{g}\Pi(g)_{ii}^{-1}$.  Therefore, it is reasonable to infer that one cannot define the same initial (or final) state in the handlebody spacetimes $\Pi(g)$ and $-\Pi(g)^{-1}$. In our model, the initial state is specified as the spatial slice at the initial time of the parent universe  $\lim\limits_{T\rightarrow 0}$TAdS$_{3}$. Assuming that this initial state can be defined in the spacetime $\Pi(g)$, then in the spacetime $-\Pi(g)^{-1}$, one can not define the same initial state. Therefore, the spacetime $-\Pi(g)^{-1}$ should not be included in the integral domain $\mathfrak{F}(g)$. Consequently, specifying the initial state breaks the modular symmetry $\mathrm{Sp}(2g, \mathbb{Z})$.

We denote the group freely generated by the generator $G(g)^{(1)}$ as $\mathbf{P}$. The group $\mathbf{P}$ is a subgroup of  $\mathrm{Sp}(2g, \mathbb{Z})$.  One can easily verify that the modular transformations  in the group $\mathbf{P}$ act trivially on the tunneling action $B_{out}$ and the vacuum decay seed $B_{seed}$. Therefore, in our model, the spacetimes associated by modular transformations in the group $\mathbf{P}$ are physically equivalent. In studies of the partition function of high genus AdS$_{3}$ spacetimes, other researchers have also suggested that the spacetimes associated with the group $\mathbf{P}$ are equivalent~\cite{XY1,AE2,SA,JM}. Considering this, when summing over different BWFs, one does not need to account for the spacetimes connected by modular transformations which are generated by $G(g)^{(1)}$ and $G(g)^{(2)}$. Since any element of the modular group $\mathrm{Sp}(2g, \mathbb{Z})$ can be generated by $G(g)^{(1)}$ and $G(g)^{(2)}$, modular transformations in the modular group $\mathrm{Sp}(2g, \mathbb{Z})$ do not need to be included in this summation.

The fundamental domain of the modular group $\mathrm{Sp}(2g, \mathbb{Z})$ in the space $\mathbf{Sie}(g)$ can be expressed as $\mathbf{Sie}(g)/\mathrm{Sp}(2g, \mathbb{Z})$. Any point in $\mathbf{Sie}(g)/\mathrm{Sp}(2g, \mathbb{Z})$ cannot be associated with another point by transformations in the modular group $\mathrm{Sp}(2g, \mathbb{Z})$. Moreover, any point in the space $\mathbf{Sie}(g)$ can be mapped to a single point in $\mathbf{Sie}(g)/\mathrm{Sp}(2g, \mathbb{Z})$. The fundamental domain $\mathbf{Sie}(g)/\mathrm{Sp}(2g, \mathbb{Z})$ is a connected subspace of $\mathbf{Sie}(g)$.
Any period matrix outside the fundamental domain $\mathbf{Sie}(g)/\mathrm{Sp}(2g, \mathbb{Z})$ can always be associated with  a unique period matrix inside it. Thus, when summing over different BWFs, one only needs to consider the period matrices within the fundamental domain $\mathbf{Sie}(g)/\mathrm{Sp}(2g, \mathbb{Z})$. In other words, spacetimes outside the fundamental domain $\mathbf{Sie}(g)/\mathrm{Sp}(2g, \mathbb{Z})$ should not be included in the integral domain $\mathfrak{F}(g)$.  Furthermore,  handlebodies within the fundamental domain $\mathbf{Sie}(g)/\mathrm{Sp}(2g, \mathbb{Z})$ that do not satisfy condition  \eqref{eq:3.45} should also be excluded from the integral domain $\mathfrak{F}(g)$. Therefore, the integral domain $\mathfrak{F}(g)$ is a subspace of the fundamental domain $\mathbf{Sie}(g)/\mathrm{Sp}(2g, \mathbb{Z})$.

\begin{figure}[tbp]
\centering
\includegraphics[width=8cm]{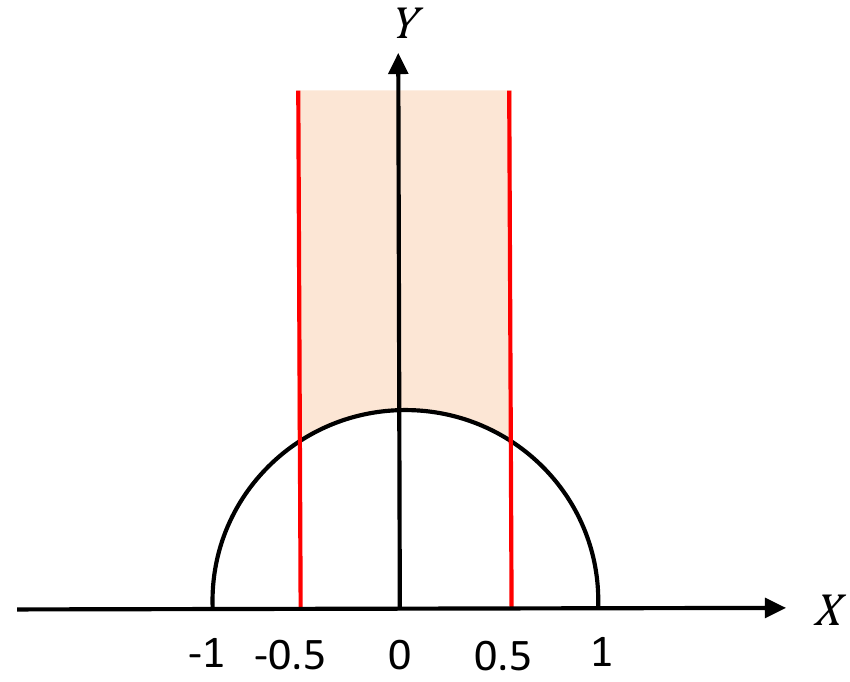}
\caption{\label{fig:6}  The fundamental domain of the modular group $\mathrm{Sp}(2, \mathbb{Z})$ in the Siegel upper half space $\mathbf{Sie}(1)$.   The shaded region represents the fundamental domain $\mathbf{Sie}(1)/\mathrm{Sp}(2, \mathbb{Z})$. $X$ and $Y$ represent the real part and imaginary part of the period matrix $\Pi(1)$, respectively.  The equations of the red lines are $X=\pm 0.5$, and the equation of the black curve is $X^{2}+Y^{2}=1$ with $Y>0$. The intersections of the black curve and the red lines are $(\pm 0.5, \frac{\sqrt{3}}{2})$. }
\end{figure}

It is well known that the fundamental domain of the modular group $\mathrm{Sp}(2, \mathbb{Z})$ in the space $\mathbf{Sie}(1)$ is determined by the equation~\cite{NK}
\begin{eqnarray}
\label{eq:4.7}
\mathbf{Sie}(1)/\mathrm{Sp}(2, \mathbb{Z})=
 \begin{cases}
      | \Pi(1)| \geq 1 \\
      | X|=| \mathrm{Re}(\Pi(1))|\leq \frac{1}{2}
  \end{cases}.
\end{eqnarray}
Here, $\mathrm{Re}(\Pi(1))$ represents the real part of the period matrix $\Pi(1)$. The shaded region in figure~\ref{fig:6} represents the fundamental domain $\mathbf{Sie}(1)/\mathrm{Sp}(2, \mathbb{Z})$. It is a two dimensional connected space. No two points within the shaded region (excluding boundary points) can be associated by modular transformation in the group $\mathrm{Sp}(2, \mathbb{Z})$. Consequently, different points in the shaded region correspond to different handlebody spacetimes, and each point satisfies equation \eqref{eq:4.7}. The boundaries of the fundamental domain $\mathbf{Sie}(1)/\mathrm{Sp}(2, \mathbb{Z})$ are defined by the equations $X=\pm 0.5$ and $X^{2}+Y^{2}=1$ with $Y>0$.

In the case where the BWFs produce only one wormhole, the period matrix is given by
\begin{equation}
\label{eq:4.8a}
\Pi(2)=
\begin{pmatrix}
\Pi(2)_{11} & \Pi(2)_{12}\\
\Pi(2)_{21} & \Pi(2)_{22}\\
\end{pmatrix}.
\end{equation}
Here, $\Pi(2)_{11}$ and $\Pi(2)_{22}$ correspond to the parent universe $\lim\limits_{T\rightarrow 0}$TAdS$_{3}$ and the wormhole, respectively. The term $\Pi(2)_{12}=\Pi(2)_{21}$ represents the interaction between the parent universe and the wormhole. The temperature of the parent universe being zero implies that $\Pi(2)_{11}=\infty$.  When the interaction $\Pi(2)_{12}=0$, the period matrix simplifies to $\Pi(2)=\Pi(2)_{11}\bigoplus\Pi(2)_{22}$, indicating that the parent universe and the wormhole are disconnected. In this case,  the wormhole degenerates into a genus one universe (a solid torus).  Various points in the shaded region of figure~\ref{fig:6} represent different genus one universes.

When there is a small interaction between the parent universe and the wormhole, the period matrix \eqref{eq:4.8a} can be expressed as $\Pi(2)=\Pi(2)_{11}\bigoplus\Pi(2)_{22} +o(\Pi(2)_{12})$, where $o(\Pi(2)_{12})$ is a small term determined by the interaction $\Pi_{12}(2)$. The predominant part of the period matrix $\Pi(2)$ is $\Pi(2)_{11}\bigoplus\Pi(2)_{22} $.  Since the interaction is small, its impact on the properties of the period matrix $\Pi(2)$  is small. Therefore, it can be anticipated  that within the fundamental domain $\mathbf{Sie}(2)/\mathrm{Sp}(4, \mathbb{Z})$, the value range of the parameter $\Pi(2)_{22}$ can be expressed as
\begin{eqnarray}
\label{eq:4.8b}
 \begin{cases}
      | \Pi(2)_{22}| \geq 1 +\delta_{1} \\
       |\mathrm{Re}(\Pi(2)_{22})|\leq \frac{1}{2}+\delta_{2}
  \end{cases}.
\end{eqnarray}
Here, $\delta_{1}$ and $\delta_{2}$ represent the influence of the interaction $\Pi(2)_{12}$. For any genus $g$, the fundamental domain $\mathbf{Sie}(g)/\mathrm{Sp}(2g, \mathbb{Z})$ satisfies the condition $|X_{ij}|\leq \frac{1}{2}$~\cite{HM}, which implies $\delta_{2}=0$.
Since $o(\Pi(2)_{12})$ is small, one can expect that $\delta_{1}$ is also small.  Therefore, equation \eqref{eq:4.7} can still approximately represent the value range of the parameter $\Pi(2)_{22}$ within the fundamental domain $\mathbf{Sie}(2)/\mathrm{Sp}(4, \mathbb{Z})$.

A small $\Pi(2)_{12}$  implies  that both $ X_{12}$ and  $Y_{12}$ are small. Denoting the upper bounds of $X_{12}$ and $Y_{12}$ as $\alpha$ and $\beta$ (meaning that if $X_{12}>\alpha$ or $Y_{12}>\beta$, condition \eqref{eq:3.45} will break down), respectively.   Then the ranges of $X_{12}$ and $Y_{12}$ are $|X_{12}|<\alpha$ ($\alpha>0$) and $0\leq Y_{12} \leq \beta$, respectively. Without the restriction of condition \eqref{eq:3.45}, $\alpha=\frac{1}{2}$~\cite{HM}. The minimum value of $Y_{12}$ being greater than zero arises from the non-negativity of the imaginary part of any element in the space $\mathbf{Sie}(g)$.
For convenience, we define $\mathfrak{D}\equiv\{(x,y) \ | \  -\alpha<x<\alpha; \ 0\leq y\leq \beta\}$. Based on these conclusions,   when $\Pi(2)_{12}$ is small, the integral domain $\mathfrak{F}(2)$ should be selected as $\mathfrak{F}(2)=\big(\mathbf{Sie}(1)/\mathrm{Sp}(2, \mathbb{Z})\big)\bigotimes \mathfrak{D}$. Points in $\mathbf{Sie}(1)/\mathrm{Sp}(2, \mathbb{Z})$ represent various boundary wormholes, while  points in $\mathfrak{D}$ represent different couplings between the wormhole and the parent universe.

Under the dilute wormhole gas approximation, different boundary wormholes are independent of each other. Each wormhole corresponds to an integral domain  $\mathfrak{F}(2)$.  Therefore, the integral domain  $\mathfrak{F}(g)$ should be
\begin{equation}
\label{eq:4.8c}
\mathfrak{F}(g)=\bigotimes\limits_{j=2}^{g}\mathfrak{F}(2)=\bigotimes\limits_{j=2}^{g}\Big(\big(\mathbf{Sie}(1)/\mathrm{Sp}(2, \mathbb{Z})\big)\bigotimes\mathfrak{D}\Big).
\end{equation}
Combining equations \eqref{eq:4.7} and \eqref{eq:4.8c}, one can show that equation \eqref{eq:4.8c} can be written as
\begin{eqnarray}
\label{eq:4.8d}
\mathfrak{F}(g)=
 \begin{cases}
 |X_{jj}|\leq \frac{1}{2}\\
      Y_{jj}\geq (1-X_{jj}^{2})^{\frac{1}{2}} \\
      |X_{1j}|\leq \alpha\\
     0\leq Y_{1j} \leq \beta
  \end{cases},\ \ \ \ \ \ j=2, 3, \cdot\cdot\cdot, g.
\end{eqnarray}
The parameters $\alpha$ and $\beta$ are subject to two conditions. The first condition is that equation \eqref{eq:3.45} must not be violated. The second condition is that the integral domain $\mathfrak{F}(g)$ must be a subspace of the fundamental domain $\mathbf{Sie}(g)/\mathrm{Sp}(2g, \mathbb{Z})$.

Equation \eqref{eq:4.8d} is derived under condition \eqref{eq:3.45}. Utilizing equations \eqref{eq:3.45} and \eqref{eq:2.16} (refer to appendix~\ref{sec:A1}), one can show that
\begin{equation}
\label{eq:4.9}
|k_{j}|=\mathrm{exp}(-2\pi Y_{jj})\leq 0.0043.
\end{equation}
Thus, under condition \eqref{eq:3.45},  it is sufficient to approximate the Schottky group $\mathbf{Sch}(L_{1}, L_{2}, ..., L_{g} )$ to the first order of the multiplier $k_{j}$.

For the integral measure $d\sigma(g)$, equation \eqref{eq:4.8d} implies that the integral variables are $(X_{22}, X_{33}, \cdot\cdot\cdot, X_{gg}, Y_{22}, Y_{33}, \cdot\cdot\cdot, Y_{gg}, X_{12}, X_{13}, \cdot\cdot\cdot, X_{1g}, Y_{12}, Y_{13}, \cdot\cdot\cdot, Y_{1g})$. Thus, the most general form of $d\sigma(g)$ is
\begin{eqnarray}\begin{split}
\label{eq:4.10}
d\sigma(g)=&\mathfrak{G}(X_{22}, ..., Y_{1g})dX_{22}dX_{33}\cdot\cdot\cdot dX_{gg}dY_{22}dY_{33}\cdot\cdot\cdot dY_{gg}\\&\times dX_{12}dX_{13}\cdot\cdot\cdot dX_{1g}dY_{12}dY_{13}\cdot\cdot\cdot dY_{1g}.
\end{split}
\end{eqnarray}
Here, $\mathfrak{G}(X_{22}, ..., Y_{1g})$ represents the determinant of the metric defined in the space $(X_{22}, X_{33}, \cdot\cdot\cdot, X_{gg}, Y_{22}, Y_{33}, \cdot\cdot\cdot, Y_{gg}, X_{12}, X_{13}, \cdot\cdot\cdot, X_{1g}, Y_{12}, Y_{13}, \cdot\cdot\cdot, Y_{1g})$.
The path integral typically corresponds to the Lebesgue measure in the configuration space. Following this reasoning, one may consider setting $\mathfrak{G}(X_{22}, ..., Y_{1g})=1$. As a result, the integral measure \eqref{eq:4.10} becomes
\begin{eqnarray}\begin{split}
\label{eq:4.10a}
d\sigma(g)=&dX_{22}dX_{33}\cdot\cdot\cdot dX_{gg}dY_{22}dY_{33}\cdot\cdot\cdot dY_{gg}\\&\times dX_{12}dX_{13}\cdot\cdot\cdot dX_{1g}dY_{12}dY_{13}\cdot\cdot\cdot dY_{1g}.
\end{split}
\end{eqnarray}
The integral measure \eqref{eq:4.10a} is the simplest one. Equation \eqref{eq:4.10a} implies that  different  handlebodies correspond to the same amplitude in the real-time path integral.

In the space $\mathbf{Sie}(g)$, the modular invariant uniquely determines the integral measure up to a constant factor~\cite{HM}. If the theory is modular invariant, the modular invariant integral measure should be used.  However,  since our model lacks modular invariance, one should not determine the function $\mathfrak{G}(X_{22}, ..., Y_{1g})$ based on modular symmetry. Additionally, the primary qualitative results of this study are independent of the specific form of the integral measure (as will be demonstrated later).  Therefore, for convenience, we choose the simplest integral measure given by equation \eqref{eq:4.10a}.

Equation \eqref{eq:4.10} and \eqref{eq:4.10a} show that the dimension of the integral space $(X_{22}, X_{33}, \cdot\cdot\cdot, X_{gg}, Y_{22}, Y_{33}, \cdot\cdot\cdot, Y_{gg}, X_{12}, X_{13}, \cdot\cdot\cdot, X_{1g}, Y_{12}, Y_{13}, \cdot\cdot\cdot, Y_{1g})$ is $4g-4$. Thus, the dimension of the integral space differs from that of the dimension of the Schottky space and the Teichm$\ddot{\mathrm{u}}$ller space of the Riemann surface $\mathbf{\Sigma}_{g}$ (refer to appendix~\ref{sec:A0}). This result arises from two main reasons. Firstly,  we only focus on BWFs that satisfy condition \eqref{eq:3.45}. Any spacetime configurations failing to meet this criterion fall outside the scope of our analysis. Secondly,  the parameters $X_{11}$ and $Y_{11}$ are associated with the parent universe. When summing over the BWFs, these parameters should not serve as integral variables.

We pointed out that without restricting the interactions between the parent universe and the wormholes to be small, equations \eqref{eq:3.33}-\eqref{eq:3.36} indicate that the tunneling action $B_{out}$ becomes a function of the variables $(X_{22}, X_{33}, \cdot\cdot\cdot, X_{gg}, Y_{22}, Y_{33}, \cdot\cdot\cdot, Y_{gg}, X_{12}, X_{13}, \cdot\cdot\cdot, X_{1g},Y_{12}, Y_{13}, \cdot\cdot\cdot, Y_{1g}, X_{23}, X_{24}, \cdot\cdot\cdot, X_{2g}, Y_{23}, Y_{24}, \cdot\cdot\cdot, Y_{2g})$. In such a scenario, the dimension of the integral space increases to $6g-8$. Thus, after subtracting the dimensions related to the multiplier $k_{1}$, the dimension of the Schottky space of the Riemann surface $\mathbf{\Sigma}_{g}$  matches that of the integral space.

To sum up,  we show that specifying the initial state  breaks the modular symmetry of the model. The integral domain $\mathfrak{F}(g)$ is a subspace of the fundamental domain $\mathbf{Sie}(g)/\mathrm{Sp}(2g, \mathbb{Z})$. Under condition \eqref{eq:3.45}, the integral domain $\mathfrak{F}(g)$  can be selected as \eqref{eq:4.8d}. Using equations \eqref{eq:4.1}, \eqref{eq:4.8d}, and \eqref{eq:4.10a}, one can complete the summation over physically non-equivalent BWFs. Although the parameters $\alpha$ and  $\beta$ in equation \eqref{eq:4.8d} remain  unspecified, we will show that their precise values are  irrelevant to the qualitative results.

\section{The tunneling rate}
\label{sec:5}
\subsection{Boundary wormhole creation rate}
\label{sec:5.1}

Generally, up to the leading order of the WKB approximation, the quantum tunneling rate is equal to the tunneling probability~\cite{SFV,EJW,HY}. Thus, in the semiclassical region, the tunneling rate from the initial state $|i\rangle$ to the final state $|f\rangle$ is given by
\begin{equation}
\label{eq:5.1}
\mathrm{exp}\big\{-2\big(S_{E}(|i\rangle\rightarrow |f\rangle)-S_{E}(|i\rangle)\big)\big\}.
\end{equation}
Here, $S_{E}(|i\rangle\rightarrow |f\rangle)$ represents the Euclidean action of the instanton trajectory $|i\rangle\rightarrow |f\rangle$.  The instanton trajectory is the solution to the Euclidean Euler-Lagrange dynamical equation. And $S_{E}(|i\rangle)$ represents the Euclidean action of the process in which the system remains in the initial state $|i\rangle$.  One can check that in the Coleman-De Luccia theory, the factor $e^{-B_{seed}}$ exhibits a similar structure to equation \eqref{eq:5.1}.

Comparing equations \eqref{eq:3.13}, \eqref{eq:3.17}, \eqref{eq:3.20}, and \eqref{eq:5.1}, it can be observed that the factor $e^{-B_{out}}$ has a similar structure to equation \eqref{eq:5.1}.  The factor $\lim\limits_{T\rightarrow 0}V(\phi_{2}, T) Rvol_{T}(g=n+1)$ in equation \eqref{eq:3.20} represents the regularized Euclidean action of the process in which the BWFs produce $g-1$ wormholes.
Therefore, the factor $e^{-B_{out}}$
should be interpreted as the rate to create the genus $g-1$ configuration $(\Pi(g)_{22}, \Pi(g)_{33}, \cdot\cdot\cdot, \Pi(g)_{gg}, \Pi(g)_{12}, \Pi(g)_{13}, \cdot\cdot\cdot, \Pi(g)_{1g})$. Consequently, the total tunneling rate  from the genus zero AdS$_{3}$ state to the genus $g-1$ AdS$_{3}$  state is
\begin{equation}
\label{eq:5.2}
\Gamma(g-1)=\int_{\mathfrak{F}(g)}e^{-B_{out}}d\sigma(g).
\end{equation}
The integral domain $\mathfrak{F}(g)$ and the tunneling action $B_{out}$ are given by equations \eqref{eq:4.8d} and \eqref{eq:3.46}, respectively. The integral measure $d\sigma(g)$ is given by equation \eqref{eq:4.10a}.

In equation \eqref{eq:5.2}, various boundary wormholes (including both microscopic  and macroscopic wormholes) are integrated.  However, equation \eqref{eq:3.17} shows that a larger wormhole volume corresponds to a larger  tunneling action $B_{out}$. Thus, the tunneling rate $e^{-B_{out}}$ for a larger boundary wormhole is smaller, indicating that larger boundary wormholes are more difficult to create. Consequently, microscopic BWFs are dominant.

In our model, $\Pi(2)_{11}=\infty$. Thus, the genus one spacetime can be parameterized by the parameters $(\Pi(2)_{22}, \Pi(2)_{12})$. This type of spacetime represents a fluctuation process of a boundary wormhole.   Setting $g=2$ in equation \eqref{eq:3.46}, one can obtain that the rate of creating the configuration $(\Pi(2)_{22}, \Pi(2)_{12})$:
\begin{equation}
\label{eq:5.3}
\Gamma(\Pi(2)_{22}, \Pi(2)_{12})=\mathrm{exp}\big\{-2\pi\Xi(\phi_{2})Y_{22}\big\}(2\pi Y_{12})^{4\Xi(\phi_{2})}.
\end{equation}
Equation \eqref{eq:5.3} shows that the rate $\Gamma(\Pi(2)_{22}, \Pi(2)_{12})$ depends on the parameters $V(\phi_{2})$, $Y_{22}$ and $Y_{12}$.
$V(\phi_{2})$ is the cosmological constant, and the parameter $Y_{12}$ represents the coupling strength between the parent universe and the wormhole.

\begin{figure}[tbp]
\centering
\includegraphics[width=14cm]{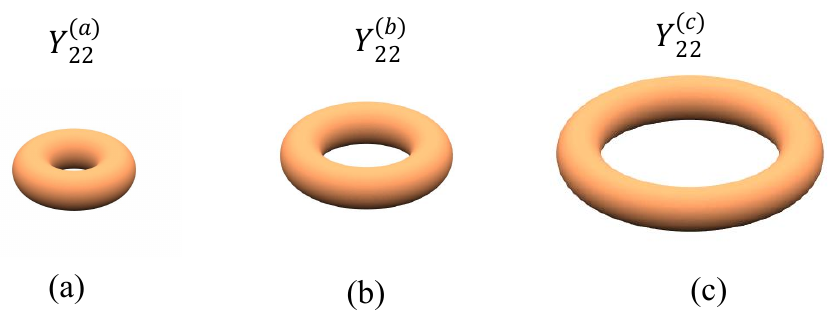}
\caption{\label{fig:6a} Three conformal nonequivalent tori.  The parameters $Y_{22}^{(a)}$, $Y_{22}^{(b)}$, and $Y_{22}^{(c)}$ represent the imaginary part of the modular parameter of the torus (a), (b), and (c), respectively.  We have set $Y_{22}^{(c)}>Y_{22}^{(b)}>Y_{22}^{(a)}>1$. Torus (a) is smaller than torus (b), which in turn is smaller than torus (c).}
\end{figure}

To illustrate the meaning of the parameter $Y_{22}$, we consider the limit as $Y_{12}$ approaches zero. In this scenario, the wormhole becomes a torus $\Pi(2)_{22}$  and is disconnected from the parent universe.  The lattice $L(1, \Pi(2)_{22})\equiv \{m+n\Pi(2)_{22}  \mid m, n \in \mathbb{Z}\} $ in the complex plane $\mathbb{C}$ can be used to represent the torus~\cite{MN}. By connecting the points $(0, 0)$, $(1, 0)$, $(\mathrm{Re}(\Pi(2)_{22}), \mathrm{Im}(\Pi(2)_{22}))$, $(\mathrm{Re}(\Pi(2)_{22})+1, \mathrm{Im}(\Pi(2)_{22}))$, one obtains a parallelogram in the complex plane. The base and height of this parallelogram are $1$ and $\mathrm{Im}(\Pi(2)_{22})=Y_{22}$, respectively. By gluing the parallel edges of this parallelogram together, one obtains a torus~\cite{MN}. Equation \eqref{eq:5.3} shows that the creation rate $\Gamma(\Pi(2)_{22}, \Pi_{12})$ is independent of the parameter $X_{22}$. For simplicity, we consider the case where $X_{22}=0$. In this instance, equation \eqref{eq:4.8d} indicates that $Y_{22}\geq 1$.  The base of the parallelogram is fixed to 1. Thus, a smaller height $Y_{22}$ corresponds to a smaller torus, as shown in figure~\ref{fig:6a}. Observing figure~\ref{fig:6a}, it can be seen that a torus with a smaller $Y_{22}$ appears ``fatter."  When $Y_{12}\neq 0$, this torus connects with the parent universe and becomes a wormhole. Therefore, a smaller $Y_{22}$ corresponds to a  smaller wormhole.

\begin{figure}[tbp]
\centering
\includegraphics[width=14cm]{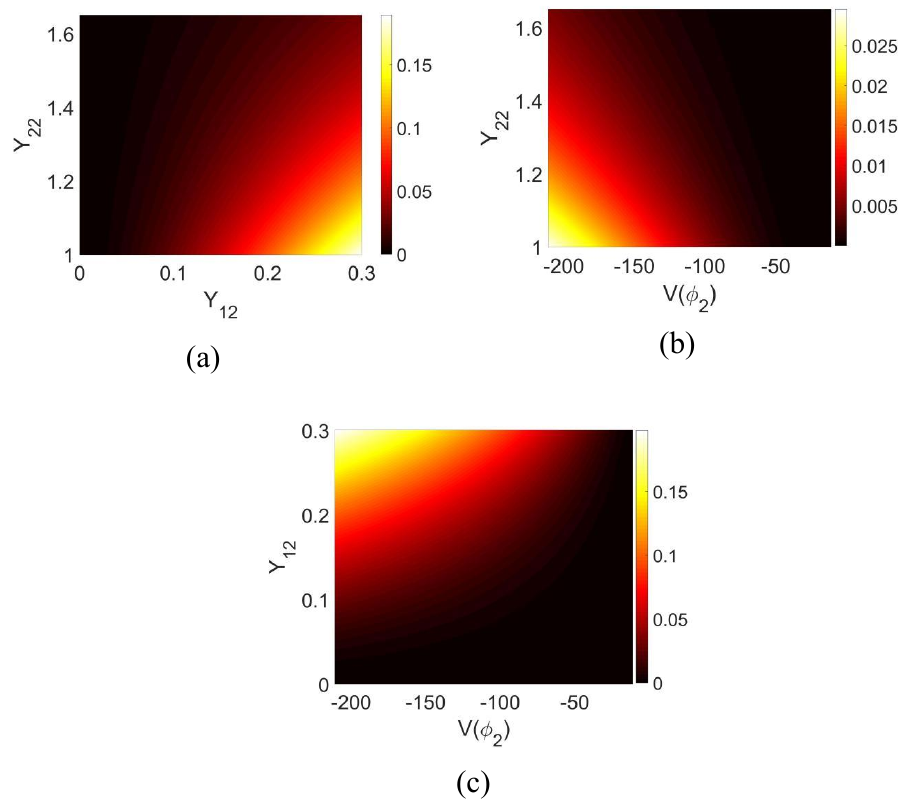}
\caption{\label{fig:7} Variations in the boundary wormhole creation rate $\Gamma(\Pi(2)_{22}, \Pi_{12})$.  In figures (a) and (b), the vertical axes represent the parameter $Y_{22}$. In figures (b) and (c), the horizontal axes represent the cosmological constant $V(\phi_{2})$. The horizontal axis of figure (a) and the vertical axis of figure (b) represent the coupling strength $Y_{12}$. Different colors in figures (a), (b) and (c) represent various values of the boundary wormhole creation rate $\Gamma(\Pi(2)_{22}, \Pi_{12})$.  The parameters are set as follows: in figure (a), $V(\phi_{2})=-200$; in figure (b), $Y_{12}=0.1$; in figure (c), $Y_{22}=1$.   }
\end{figure}

Figure~\ref{fig:7} illustrates variations in the boundary wormhole creation rate $\Gamma(\Pi(2)_{22}, \Pi_{12})$. Figures ~\ref{fig:7}(a) and ~\ref{fig:7}(c) show that stronger coupling strength corresponds to a larger boundary wormhole creation rate $\Gamma(\Pi(2)_{22}, \Pi_{12})$. Figures~\ref{fig:7}(a) and ~\ref{fig:7}(b) indicate that as the parameter $Y_{22}$ increases, the boundary wormhole creation rate $\Gamma(\Pi(2)_{22}, \Pi_{12})$ decreases.  This suggests  that smaller wormholes are easier to create in the parent universe $\lim\limits_{T\rightarrow 0}\mathrm{TAdS}_{3}$. Figures~\ref{fig:7}(b) and ~\ref{fig:7}(c) reveal that as the absolute value of the cosmological constant increases, the boundary wormhole creation rate  $\Gamma(\Pi(2)_{22}, \Pi_{12})$ also increases. This implies that BWFs are more likely to occur in the AdS$_{3}$ spacetime with a more negative cosmological constant. Equation \eqref{eq:5.3} clearly indicates that the boundary wormhole creation rate $\Gamma(\Pi(2)_{22}, \Pi_{12})$ is independent of the integral measure $d\sigma(g)$. Therefore, the conclusions drawn from figure~\ref{fig:7} are also independent of the integral measure.

For multiple BWFs, as shown in figure~\ref{fig:m1}(b), the tunneling rate from the genus zero spacetime state to the genus $g-1$ spacetime state is given by  equation \eqref{eq:5.2}. Substituting equations \eqref{eq:3.46}, \eqref{eq:4.8d}, and \eqref{eq:4.10a} into \eqref{eq:5.2},  the tunneling rate $\Gamma(g-1)$ can be written as
\begin{eqnarray}\begin{split}
\label{eq:5.4}
\Gamma(g-1)=&\int_{-\frac{1}{2}}^{\frac{1}{2}}dX_{22}\int_{-\frac{1}{2}}^{\frac{1}{2}}dX_{33}\cdot\cdot\cdot\int_{-\frac{1}{2}}^{\frac{1}{2}}dX_{gg}\int_{\sqrt{1-X_{22}^{2}}}^{\infty}dY_{22}\int_{\sqrt{1-X_{33}^{2}}}^{\infty}dY_{33}\cdot\cdot\cdot\int_{\sqrt{1-X_{gg}^{2}}}^{\infty}dY_{gg}\\&\int_{-\alpha}^{\alpha}dX_{12}\int_{-\alpha}^{\alpha}dX_{13}\cdot\cdot\cdot\int_{-\alpha}^{\alpha}dX_{1g}\int_{0}^{\beta}dY_{12}\int_{0}^{\beta}dY_{13}\cdot\cdot\cdot\int_{0}^{\beta}dY_{1g}\\&\ \ \times\prod_{j=2}^{g}\Big(\mathrm{exp}\{-2\pi \Xi(\phi_{2})Y_{jj}\}(2\pi Y_{12})^{2\Xi(\phi_{2})}(2\pi Y_{1j})^{2\Xi(\phi_{2})}  \Big).
\end{split}
\end{eqnarray}
Using the approximation $\sqrt{1-X_{jj}^{2}}\approx 1-\frac{1}{2}X_{jj}^{2} +o(X_{jj}^{4})$, one can complete the integration over the variables $(X_{22}, X_{33}, \cdot\cdot\cdot, X_{gg},Y_{22}, Y_{33}, \cdot\cdot\cdot, Y_{gg}, X_{12}, X_{13}, \cdot\cdot\cdot, X_{1g}, Y_{12}, Y_{13}, \cdot\cdot\cdot, Y_{1g})$. Consequently, equation \eqref{eq:5.4} simplifies to
\begin{equation}
\label{eq:5.5}
\Gamma(g-1)=\frac{2\Xi(\phi_{2})+1}{2g\Xi(\phi_{2})+1}\Big(\Theta(\phi_{2})\mathrm{erfi}\big(\frac{1}{2}\sqrt{\pi \Xi(\phi_{2})}\big)\Big)^{g-1},
\end{equation}
where,
\begin{eqnarray}\begin{split}
\label{eq:5.6}
\Theta(\phi_{2})\equiv 2\alpha\frac{\mathrm{exp}\{-2\pi\Xi(\phi_{2})\}}{2\pi\Xi(\phi_{2})}\frac{(2\pi)^{4\Xi(\phi_{2})}}{2\Xi(\phi_{2})+1}\Xi(\phi_{2})^{-\frac{1}{2}}\beta^{4\Xi(\phi_{2})+1}
\end{split}
\end{eqnarray}
and
\begin{equation}
\label{eq:5.7}
\mathrm{erfi}(x)\equiv \frac{2}{\sqrt{\pi}}\int_{0}^{x}e^{t^{2}}dt=\frac{2}{\sqrt{\pi}}\sum_{k=0}^{\infty}\frac{x^{2k+1}}{(2k+1)k!}
\end{equation}
is the imaginary error function.

In equation \eqref{eq:5.4}, the factor
\begin{equation}
\label{eq:5.7a}
\prod_{j=2}^{g}\Big(\mathrm{exp}\{-2\pi \Xi(\phi_{2})Y_{jj}\}(2\pi Y_{12})^{2\Xi(\phi_{2})}(2\pi Y_{1j})^{2\Xi(\phi_{2})}  \Big)
\end{equation}
is equal to $e^{-B_{out}}$. Thus, this factor  represents the tunneling probability from the parent universe $\lim\limits_{T\rightarrow 0}$TAdS$_{3}$ to the genus $g-1$ configuration $(\Pi(g)_{22}, \Pi(g)_{33}, \cdot\cdot\cdot, \Pi(g)_{gg}, \Pi(g)_{12}, \Pi(g)_{13}, \cdot\cdot\cdot, \Pi(g)_{1g})$. Up to the leading order of the WKB approximation, it also denotes the tunneling rate.

Under the dilute wormhole gas approximation, the fluctuations of each wormhole are independent of one another. Thus, the factor \eqref{eq:5.7a} can be expressed as the multiplication of the tunneling probabilities of each wormhole.
To illustrate this, we temporarily rewrite $e^{-B_{out}}$ in the Schottky space. In the small $k_{j}$ limit, utilizing equation \eqref{eq:3.26}, one can derive
\begin{equation}
\label{eq:5.7b}
e^{-B_{out}}=\prod_{j=2}^{g}|\sqrt{k_{j}}(\xi_{j}-\eta_{j})|^{2\Xi(\phi_{2})}.
\end{equation}
It is easy to show that
\begin{equation}
\label{eq:5.7c}
|\sqrt{k_{2}}(\xi_{2}-\eta_{2})|^{2\Xi(\phi_{2})}=\mathrm{exp}\big\{-2\pi\Xi(\phi_{2})Y_{22}\big\}(2\pi Y_{12})^{4\Xi(\phi_{2})}.
\end{equation}
Comparing equations \eqref{eq:5.7c} and \eqref{eq:5.3}, one finds that $|\sqrt{k_{2}}(\xi_{2}-\eta_{2})|^{2\Xi(\phi_{2})}=\Gamma(\Pi(2)_{22}, \Pi(2)_{12})$. This indicates that the factor
$|\sqrt{k_{2}}(\xi_{2}-\eta_{2})|^{2\Xi(\phi_{2})}$ signifies the tunneling probability of creating the boundary wormhole $(\Pi(2)_{22}, \Pi(2)_{12})$. Therefore, the factor $|\sqrt{k_{j}}(\xi_{j}-\eta_{j})|^{2\Xi(\phi_{2})}$ represents the tunneling probability of creating the boundary wormhole $(\Pi(2)_{jj}, \Pi(2)_{1j})$. Hence, the factor \eqref{eq:5.7a} indeed equals the product of the tunneling probabilities of each wormhole.

\begin{figure}[tbp]
\centering
\includegraphics[width=14cm]{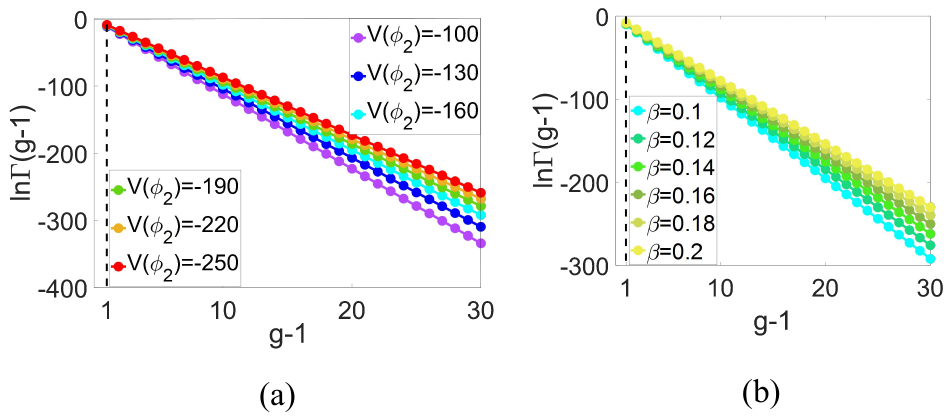}
\caption{\label{fig:8} Variations in the tunneling rate $\Gamma(g-1)$ versus genus. In figures (a) and (b), the horizontal axes represent the genus. The vertical axes represent the logarithm of the tunneling rate $\Gamma(g-1)$. The parameters are set as follows: in figure (a), $\alpha=\beta=0.1$; in figure (b), $\alpha=0.1$ and $V(\phi_{2})=-160$.}
\end{figure}

Figure~\ref{fig:8} illustrates the variations in the tunneling rate $\Gamma(g-1)$. Both figures~\ref{fig:8}(a) and ~\ref{fig:8}(b) show that as the genus increases, the tunneling rate $\Gamma(g-1)$ decreases rapidly, indicating that the dominant fluctuations are those of a single wormhole. This result can also be inferred from another perspective: under the dilute wormhole gas approximation, the fluctuations of different wormholes are independent of one another. The probability of creating multiple wormholes can be considered as the product of the probabilities of producing each wormhole. Consequently, it becomes more challenging to produce a higher genus spacetime. This leads to a decrease in $\Gamma(g-1)$ with increasing $g-1$. Following this perspective,  regardless of the integral measure $d\sigma(g)$ employed, as long as the fluctuations of different wormholes remain independent, one can expect this conclusion to be qualitatively valid.

Figure~\ref{fig:8}(a) also reveals that in the AdS$_{3}$ spacetime with a more negative cosmological constant, the tunneling rate $\Gamma(g-1)$ is larger, which  is consistent with the findings from figures~\ref{fig:7}(b) and ~\ref{fig:7}(c). Figure~\ref{fig:8}(b) shows that the exact value of the parameter $\beta$ does not affect the qualitative result. In addition, equations \eqref{eq:5.5} and \eqref{eq:5.6} indicate that $\Gamma(g-1)\propto \alpha^{g-1}$. Therefore, the qualitative results obtained from figure~\ref{fig:8} are also unaffected by the specific value of the parameter $\alpha$.

Gravitational lensing effects, quasinormal modes and other observable phenomena may help to find wormholes experimentally~\cite{NT76,SY76,JRM76,V76,SSS76,SW76,FB76,DD76,RA79,RP79}. For example, in~\cite{RP79}, the authors examined distinctive features of strong gravitational lensing by Schwarzschild black holes compared to those by wormholes. One such difference is that the relativistic images of black holes are formed outside the photon spheres, whereas those of wormholes can be formed both inside and outside the photon spheres~\cite{RP79}. Thus, although both wormholes and black holes can produce gravitational lensing effects, their characteristics may differ. These distinctive features may help distinguish wormholes from black holes in experiments.

BWFs processes lead to variations in the spacetime genus. Spacetimes with non-zero genus are multi-connected, as shown in figures~\ref{fig:m1} and ~\ref{fig:m3}. In~\cite{MJ}, the authors summarized several observable effects of a multi-connected universe. One interesting effect is the existence of ghost images (the image nearest to the observer is called the real image, while the others are referred to as ghost images), which arise from the fact that in a multi-connected universe, there are typically multiple null geodesics connecting the emitting source to the observer. Different ghost images correspond to different redshifts and look-back times. These ghost images may provide a way to detect the multi-connected structure of spacetime. More detailed discussions on this topic can be found in~\cite{MJ,DV76}.

If both wormholes and the multi-connected structure of the universe can be observed in the future, this may serve as a potential clue for the wormhole fluctuations. However, wormholes can be generated not only through wormhole fluctuations; they may also be produced by other mechanisms, such as the formation of Schwarzschild black holes or the evolution of false vacuum bubbles~\cite{SEA,WDJ77}. Therefore, such observational clues may not be sufficient to confirm the existence of wormhole fluctuations. Since BWFs are quantum processes, and current observations suggest that the present universe behaves classically, it may be difficult to directly detect BWFs through the classical cosmological observations. However, the early universe remnants might provide a clue.

According to holographic duality, an AdS-Schwarzschild wormhole is dual to a thermofield double  state of the SYK system~\cite{J712,MJ712,PDA712}. The properties of the AdS-Schwarzschild wormhole can be derived from both gravity theory and the SYK system~\cite{MJ712}.  In~\cite{DAJ77}, the authors experimentally simulated wormholes using a thermofield double state of the SYK system. The experimental results (such as variations in mutual information) are consistent with theoretical predictions.  Therefore,  beyond astronomical observations,  wormholes may also be simulated experimentally~\cite{DAJ77,CS77}. Consequently, it is natural to expect that wormhole fluctuations could also be simulated in the future. In short, detecting wormhole fluctuations remains an important open problem that deserves further investigation.

To sum up, in this section, we demonstrate that the rate of creating the spacetime configuration $(\Pi(2)_{22}, \Pi_{12})$ is described by the quantity $\Gamma(\Pi(2)_{22}, \Pi_{12})$. We show that BWFs are more likely to occur in AdS$_{3}$ spacetime with a more negative cosmological constant. We also show that smaller wormholes are easier to create. The total rate of tunneling from the genus zero AdS$_{3}$ state to a higher genus spacetime state is described by $\Gamma(g-1)$. The rate $\Gamma(g-1)$ rapidly decreases as the genus increases. Thus, the dominant fluctuations are the genus one fluctuations.

The BWFs break the global spherical symmetry of the universe, as shown in figure~\ref{fig:m1}. Up to now, the existence of wormholes has not been confirmed by experiments. In addition, observations of the cosmic microwave background (CMB) suggest that the observed universe is highly homogeneous and isotropic. In other words, the observed universe approximately exhibits global spherical symmetry. This seems to indicate that the number of large scale boundary wormholes is small compared to the number of various celestial bodies. Thus, a question arises: given that wormholes can be created via quantum tunneling, why is the number of large scale wormholes small?  Our calculations for $\Gamma(\Pi(2)_{22}, \Pi_{12})$ and $\Gamma(g-1)$ indicate that macroscopic boundary wormholes are difficult to create, and the probability of creating multiple wormholes is small. These results support the idea that the number of large scale wormholes should indeed be small. Although our calculations are limited to boundary wormholes in AdS$_{3}$ spacetime, we believe that our results provide valuable insights into this question.

\subsection{Boundary wormhole fluctuations influence on the false vacuum decay}
\label{sec:5.2}
If  wormhole fluctuations are not considered, the vacuum decay rate is given by $e^{-B_{seed}}$. However, when BWFs are taken into account, vacuum decay can occur through multiple pathways. During the decay from the false vacuum state to the true vacuum state, the spacetime may produce various wormholes. All these possible spacetime configurations must be accounted for. Therefore, when considering BWFs, the vacuum decay rate is modified as described in equation \eqref{eq:4.1}.

\begin{figure}[tbp]
\centering
\includegraphics[width=14cm]{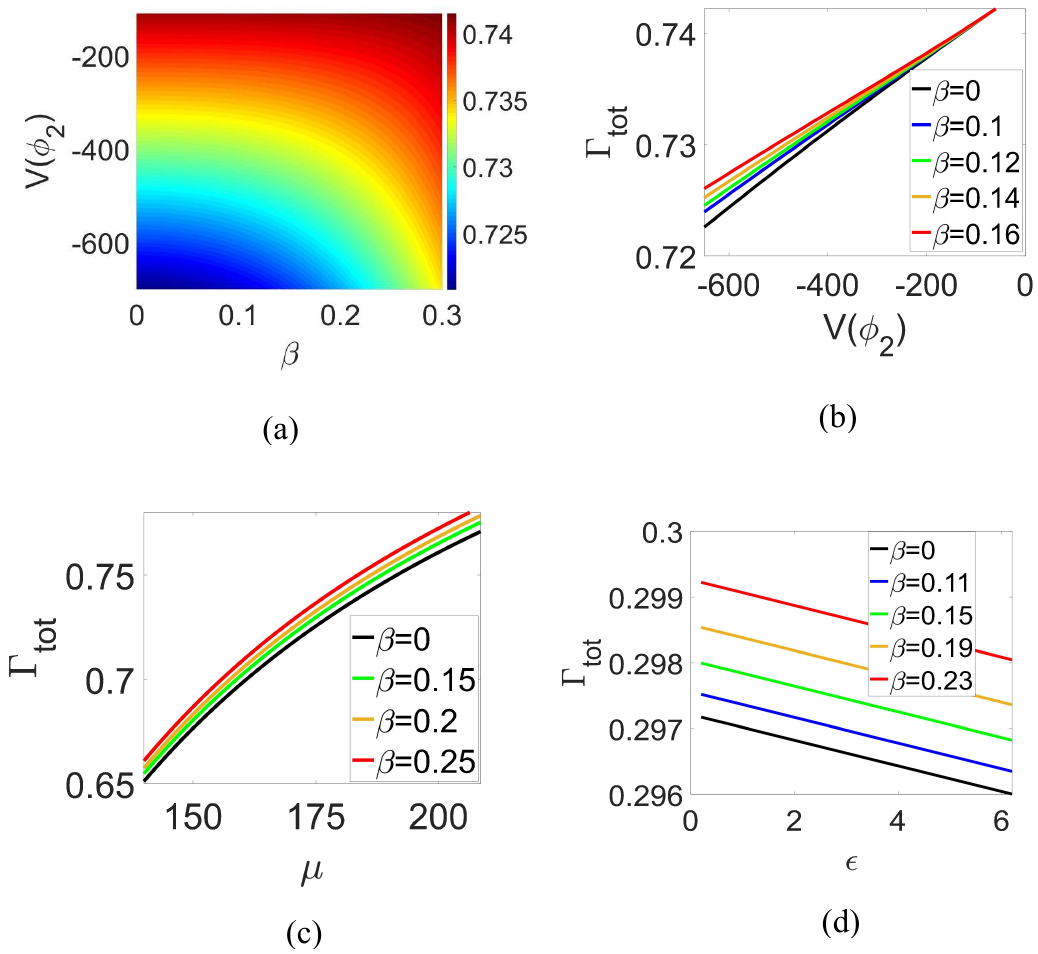}
\caption{\label{fig:9} Variations in the  vacuum decay rate $\Gamma_{tot}$ under the condition $\alpha=0.2$. In figure (a), the horizontal axis and the vertical axis represent the upper limit of the coupling strength $Y_{12}$ and the false vacuum energy density $V(\phi_{2})$, respectively. Different colors represent various values of the vacuum decay rate $\Gamma_{tot}$. In figures (b), (c) and (d), the vertical axes represent the  vacuum decay rate $\Gamma_{tot}$. The horizontal axes of figures (b), (c) and (d) represent the false vacuum energy density $V(\phi_{2})$, the tension of the domain wall and the difference of the two vacuum energy density, respectively.  Different curves correspond to different values of the parameter $\beta$. The black curves represent the vacuum decay rate in the Coleman-De Luccia theory. The parameters are set as follows: in figures (a) and (b),  $\mu=150$ and $\epsilon=0.1$; in figure (c), $V(\phi_{2})=-800$ and $\epsilon=1$; in figure (d),  $V(\phi_{2})=-500$ and $\mu=70$.}
\end{figure}

\begin{figure}[tbp]
\centering
\includegraphics[width=14cm]{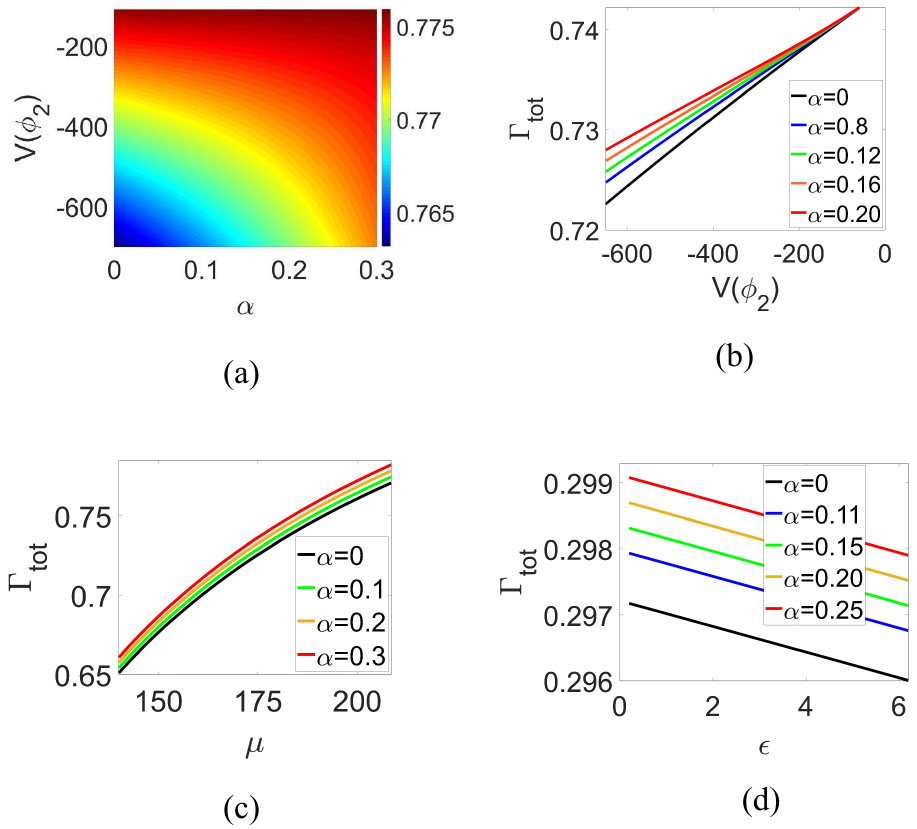}
\caption{\label{fig:10} Variations in the  vacuum decay rate $\Gamma_{tot}$ under the condition $\beta=0.2$. In figure (a), the horizontal axis and the vertical axis represent the upper limit of the coupling strength $Y_{12}$ and the false vacuum energy density $V(\phi_{2})$, respectively. Different colors represent different values of the vacuum decay rate $\Gamma_{tot}$. In figures (b), (c) and (d), the vertical axes represent the  vacuum decay rate $\Gamma_{tot}$. The horizontal axes of figures (b), (c) and (d) represent the false vacuum energy density $V(\phi_{2})$, the tension of the domain wall and the difference of the two vacuum energy density, respectively.  Different curves correspond to different values of the parameter $\alpha$. The black curves represent the vacuum decay rate in the Coleman-De Luccia theory. The parameters are set as follows: in figure (a), $\mu=200$ and $\epsilon=0.1$; in figure (b),  $\mu=170$ and $\epsilon=0.1$; in figure~\ref{fig:10} (c), $V(\phi_{2})=-800$ and $\epsilon=1$; in figure (d), $V(\phi_{2})=-500$ and $\mu=70$.}
\end{figure}

By comparing equations \eqref{eq:4.1} and \eqref{eq:5.2},  the relationship between the vacuum decay rate $\Gamma_{tot}$ and the  tunneling rate $\Gamma(g-1)$ is given by
\begin{equation}
\label{eq:5.8}
\Gamma_{tot}=e^{-B_{seed}}+e^{-B_{seed}}\sum_{g=2}^{\infty}\Gamma(g-1).
\end{equation}
Substituting equation \eqref{eq:5.5} into \eqref{eq:5.8} yields
\begin{equation}
\label{eq:5.9}
\Gamma_{tot}=\frac{2\Xi(\phi_{2})+1}{2\Xi(\phi_{2})}e^{-B_{seed}}\ \mathrm{LerchPhi}\big[\Theta(\phi_{2})\mathrm{erfi}\big(\frac{1}{2}\sqrt{\pi \Xi(\phi_{2})}\big),\  1, \ \frac{2\Xi(\phi_{2})+1}{2\Xi(\phi_{2})}\big],
\end{equation}
where
\begin{equation}
\label{eq:5.10}
\mathrm{LerchPhi}[x,s,a]\equiv\sum_{n=0}^{\infty}\frac{x^{n}}{(a+n)^{s}}
\end{equation}
is the Lerch transcendent function~\cite{LFJ}. In equation \eqref{eq:5.9}, the factor $e^{-B_{seed}}$ is the vacuum decay rate predicted by Coleman-De Luccia theory, while the other factors represent  the contribution of BWFs.

Figure~\ref{fig:9} shows the variations in the vacuum decay rate $\Gamma_{tot}$. Figures~\ref{fig:9}(a) and ~\ref{fig:9}(b)  show that as the absolute value of the false vacuum energy density decreases, the vacuum decay rate $\Gamma_{tot}$ increases. Figure~\ref{fig:9}(b) indicates that as $|V(\phi_{2})|$ decreases, different curves will intersect at one point, implying that the effects of BWFs decrease as $|V(\phi_{2})|$ decreases. This result aligns with the result from figures~\ref{fig:7}(b), ~\ref{fig:7}(c), and figure~\ref{fig:8}(a). Figure~\ref{fig:9}(b) also indicates that the vacuum decay rate $\Gamma_{tot}$ increases with $V(\phi_{2})$. In our model, a smaller $V(\phi_{2})$ (keeping other parameters fixed) may correspond to a higher potential barrier between the two vacuum states. Thus, figure~\ref{fig:9}(b) implies that a higher potential barrier corresponds to a harder tunneling. This is a typical feature of  quantum tunneling processes. Figure~\ref{fig:9}(c) reveals that as the tension of the domain wall increases, the vacuum decay rate $\Gamma_{tot}$ also increases. Furthermore, figure~\ref{fig:9}(d) shows that a smaller energy density difference between the false vacuum state and the true vacuum state corresponds to a higher vacuum decay rate.

The black curves in figures~\ref{fig:9}(b), ~\ref{fig:9}(c), and ~\ref{fig:9}(d) represent the variations in the vacuum decay rate without wormhole fluctuations. These figures indicate that including BWFs increases  the vacuum decay rate, as more pathways for vacuum decay are considered. This result is valid regardless of the integral measure $d\sigma(g)$. Figures~\ref{fig:8}(b),~\ref{fig:9}(a), ~\ref{fig:9}(b), ~\ref{fig:9}(c), and ~\ref{fig:9}(d) illustrate that the specific value of parameter $\beta$ does not affect the qualitative conclusions. Figure~\ref{fig:10} is similar to figure~\ref{fig:9}, indicating that the particular value of parameter $\alpha$ also has no influence  on the qualitative results.

\section{Conclusions and discussions}
\label{sec:6}

In this work, we studied the BWFs of the AdS$_{3}$ spacetime and their influence on false vacuum decay. Our model consists of a real scalar field coupled with the three dimensional general relativity. The scalar potential has two minimal values, corresponding to two vacuum states.
According to Coleman-De Luccia theory, the vacuum decay rate is determined by the total tunneling action, which consists of the  Euclidean action of the bounce solution and a background term. Under the thin wall approximation, the bounce solution can be visualized as a spherically  symmetric bubble embedded in a sea of false vacuum. The total tunneling action can be divided into three components: the tunneling action inside, within, and outside the domain wall.  For convenience, we define the vacuum decay seed as the sum of the tunneling action inside and on the domain wall. Without wormhole fluctuations, the tunneling action outside the bubble is zero, making the total tunneling action equivalent to the vacuum decay seed.

We demonstrate that in the presence of BWFs, the tunneling action outside the bubble becomes nonzero.  The volume of AdS$_{3}$ spacetime is infinite. Thus, studying the tunneling action outside the bubble requires regularizing the volume and action of spacetime. Both the regularized action and the original Einstein-Hilbert action correspond to the same spacetime dynamical equations.

We categorize wormhole fluctuations into three classes. In the first class, both ends of the wormhole is inside the bubble. In the second class, one end of the wormhole is inside the bubble while the other is outside. In the third class, both ends of the wormhole are outside the bubble. Fluctuations of the third class do not affect the vacuum decay seed.  Our focus is on the scenario where the bubble radius is much smaller than the size of AdS$_{3}$ spacetime. In this case, the third class of fluctuations is predominant. Thus, we disregard the first and second classes of fluctuations.

The three dimensional high genus spacetime includes handlebodies and non-handlebodies. Some researchers argue that the effects of the handlebodies are predominant. Thus, for simplicity, we focus on the contributions of the handlebodies. All three dimensional handlebodies can be obtained by filling the interior of a closed high genus Riemann surface. Therefore,  summing over different handlebodies is equivalent to summing over various closed Riemann surfaces.
There are different methods to classify the closed Riemann surfaces.  Both Schottky space and the Siegel upper half space can be used to parameterize different closed  Riemann surfaces. The summation over different closed Riemann surfaces translates into an integration in these two spaces.

The regularized action of the high genus AdS$_{3}$ spacetime in  Schottky space is helpful in deriving the tunneling action of the high genus spacetime. Thus, at first, we formulate the tunneling action in Schottky space. However, it is easier to define the integral domain in the Siegel upper half space. Therefore, we transform the tunneling action from Schottky space into the Siegel upper half space. This transformation can be simplified using the dilute wormhole gas approximation.

In the Siegel upper half space, various handlebodies are characterized by different period matrices. The off-diagonal elements are interpreted as the interactions between different wormholes or between the parent universe and the wormholes. Under the dilute wormhole gas approximation, the interactions between different wormholes are small.  However, this does not imply that the interactions between the parent universe and the wormholes are also necessarily small. To simplify the tunneling action, we assume that the interactions among different wormholes are significantly smaller  compared to the interactions between the parent universe and the wormholes.

Different handlebodies can be interpreted as various BWFs.  We show that BWFs are more likely to occur in  AdS$_{3}$ spacetime with a more negative cosmological constant. We find that smaller wormholes are easier to create.
The summation over different BWFs involves integration in the Siegel upper half space. We demonstrate that the integral domain is a subspace of the fundamental domain of the modular group $\mathrm{Sp}(2g, \mathbb{Z})$  in the Siegel upper half space. Upon completing the integration, we obtain the tunneling rate from the genus zero spacetime state to the high genus spacetime state. We find that as the genus increases, the tunneling rate decreases rapidly. Thus, the primary fluctuations are the genus one fluctuations. Taking into account the BWFs, the vacuum decay rate will increase. In our model, despite there being two unfixed parameters, we have demonstrated that the specific values of these parameters do not affect any qualitative results.

\section*{Acknowledgements}
Hong Wang  was supported by the National Natural Science Foundation of China Grant  No.12234019. Hong Wang thanks for the help from Professor Erkang Wang.

\appendix
\section{Schottky parametrization of the AdS$_{3}$ handlebody}
\label{sec:A0}

The parameterization of the AdS$_{3}$ handlebody  is essential for studying BWFs. Under a specific cosmological constant, the distinctions between various AdS$_{3}$ handlebodies are determined by their conformal boundaries. Thus, to parameterize an AdS$_{3}$ handlebody, it suffices to parameterize its conformal boundary.  The conformal boundary of any AdS$_{3}$ handlebody is a two dimensional closed Riemann surface. There are several methods for parameterizing the two dimensional closed Riemann surface. Different parameterization methods may correspond to different ways of constructing the closed Riemann surface. We focus on the Schottky parameterization. The mathematics reviewed in this section can be found in references~\cite{BM,PFM}.

The Schottky group is a discrete subgroup of the two dimensional special complex linear transformation group $SL(2,\mathbb{C})$. The elements of the Schottky group can be expressed as
\begin{equation}
\label{eq:2.1}
\mathscr{T}=
\begin{pmatrix}
a & b\\

c & d
\end{pmatrix}
,
\end{equation}
where $a$, $b$, $c$ and $d$ are complex numbers satisfying $ad-bc=1$. The Riemann sphere $\mathbb{\mathbb{S}}$ is a two dimensional complex sphere.
The operation of the Schottky group on the Riemann  sphere can be written as
\begin{equation}
\label{eq:2.2}
\mathscr{T}(z)=\frac{az+b}{cz+d}.
\end{equation}
Here, ``$z$'' represents a point on the Riemann sphere. We use the symbol $\mathbf{Sch}(L_{1}, L_{2}, ..., L_{g} )$ to represent the genus g Schottky group, where $L_{1}, L_{2}, ..., L_{g}$ represent its generators. $\mathbf{Sch}(L_{1}, L_{2}, ..., L_{g} )$ is freely generated by $L_{1}, L_{2}, ..., L_{g}$. That is, any element of $\mathbf{Sch}(L_{1}, L_{2}, ..., L_{g} )$ can be written in the form $L_{r_{1}}^{n_{1}}L_{r_{2}}^{n_{2}}...L_{r_{m}}^{n_{m}}$, where $r_{i}=1, 2, ..., g$ and $n_{i}=\pm 1, \pm2,...$.

Any genus $g$ closed Riemann surface can be constructed by the operation of the genus $g$ Schottky group on the Riemann sphere~\cite{KK,XY1,BM,LF}. The operation of a generator $L_{i}$ of the genus $g$ Schottky group on the Riemann sphere can always be expressed as~\cite{LF,XY1,BM}
\begin{equation}
\label{eq:2.3}
\frac{L_{i}(z)-\xi_{i}}{L_{i}(z)-\eta_{i}}=k_{i}\frac{z-\xi_{i}}{z-\eta_{i}}.
\end{equation}
Here, $i=1, 2, ..., g$.  $\xi_{i}$ and $\eta_{i}$ are the attractive  and the repelling fixed point of the generator $L_{i}$, respectively, and $k_{i}$ is the multiplier with $|k_{i}|<1$. For a normalized Schottky group, $\xi_{1}=0$, $\xi_{2}=1$, and $ \eta_{1}=\infty$. For a marked Schottky group, the generators $L_{1}, L_{2}, ..., L_{g}$ are ordered. Both normalized and marked conditions are necessary to parameterize the two dimensional closed Riemann surface.
Henceforth, when referring to the Schottky group, we assume it is both normalized and marked.

Equation \eqref{eq:2.3} indicates that $L_{i}(z)$ can be written as
\begin{equation}
\label{eq:2.3a}
L_{i}(z)=\frac{\eta_{i}(z-\xi_{i})-k_{i}\xi_{i}(z-\eta_{i})}{(z-\xi_{i})-k_{i}(z-\eta_{i})}.
\end{equation}
By comparing equations \eqref{eq:2.1}, \eqref{eq:2.2}, and \eqref{eq:2.3a}, one can read off the matrix representation of the generator $L_{i}$ as
\begin{equation}
\label{eq:2.3b}
L_{i}=\frac{1}{\sqrt{|k_{i}|}|\eta_{i}-\xi_{i}|}
\begin{pmatrix}
\xi_{i}-k_{i}\eta_{i} & -\eta_{i}\xi_{i}(1-k_{i})\\

1-k_{i} & k_{i}\xi_{i}-\eta_{i}
\end{pmatrix}
.
\end{equation}
According to equation \eqref{eq:2.3b}, one can obtain the matrix representation of any element in the Schottky group $\mathbf{Sch}(L_{1}, L_{2}, ..., L_{g} )$.

Denoting the multiplier of the element $L_{i}L_{j}$ as $k_{ij}$,  the relationship between $k_{ij}$ and the parameters $(\xi_{i}, \eta_{i}, k_{i}, \xi_{j}, \eta_{j}, k_{j})$ is given by~\cite{BM}
\begin{equation}
\label{eq:2.5}
k_{ij}=\frac{\mathrm{det}(L_{i}L_{j})}{(\mathrm{Tr}(L_{i}L_{j}))^{2}}+o(k_{i}^{2}, k_{j}^{2})\approx k_{i}k_{j}\frac{(\eta_{i}-\xi_{i})^{2}(\eta_{j}-\xi_{j})^{2}}{(\eta_{j}-\xi_{i})^{2}(\eta_{i}-\xi_{j})^{2}}.
\end{equation}
Here, $\mathrm{det}(L_{i}L_{j})$ and $\mathrm{Tr}(L_{i}L_{j})$ represent the determinant and the trace of the element $L_{i}L_{j}$, respectively, and  $o(k_{i}^{2}, k_{j}^{2})$ represents higher order terms of the multipliers $k_{i}$ and $k_{j}$. It is worth noting that $L_{i}(z)$ and $L_{i}^{-1}(z)$ share the same multiplier. Thus, according to equation \eqref{eq:2.5}, one can prove that the multipliers of the elements  $L_{i}L_{j}$,  $L_{j}L_{i}$,  $L_{i}^{-1}L_{j}^{-1}$, and  $L_{j}^{-1}L_{i}^{-1}$ are equivalent. Similarly, the multipliers of the elements  $L_{i}^{-1}L_{j}$,  $L_{i}L_{j}^{-1}$,  $L_{j}^{-1}L_{i}$, and  $L_{j}L_{i}^{-1}$ are also equivalent~\cite{BM}. Therefore,
up to the first order of the multipliers, the Schottky group $\mathbf{Sch}(L_{1}, L_{2}, ..., L_{g} )$ can be approximated as
\begin{equation}
\label{eq:2.6}
\mathbf{Sch}(L_{1}, L_{2}, ..., L_{g} )\approx \{id,\ L_{i}, \ L_{i}^{-1} \quad |\  i=1, 2, ..., g\}.
\end{equation}
Here, $id$ represents the identity element of the Schottky group. The second order approximation is
\begin{eqnarray}\begin{split}
\label{eq:2.7}
\mathbf{Sch}(L_{1}, L_{2}, ..., L_{g} )\approx & \{id,\ L_{i}, \ L_{i}^{-1},\ L_{i}L_{j}, \ L_{i}L_{j}^{-1}, \ L_{i}^{-1}L_{j}, \ L_{i}^{-1}L_{j}^{-1}, \\& \quad L_{i}^{2}, \ (L_{i}^{-1})^{2} \ |\ i, j=1, 2, ..., g; \ i\neq j \}.
\end{split}
\end{eqnarray}

Equation \eqref{eq:2.3} indicates that $L_{i}$ is determined by the parameters $\xi_{i}, \eta_{i}$, and $ k_{i}$. The genus $g$ Schottky group $\mathbf{Sch}(L_{1}, L_{2}, ..., L_{g} )$ is determined by its generators $L_{1}, L_{2}, ..., L_{g}$. In addition, the normalized conditions of the Schottky group are $\xi_{1}=0, \xi_{2}=1$ and $ \eta_{1}=\infty$. Thus, a genus $g$  Schottky group  is determined by the set of parameters $(k_{1}, \eta_{2}, k_{2}, \xi_{3}, \eta_{3}, k_{3}, ..., \xi_{g}, \eta_{g}, k_{g})$. For a genus one Schottky group $\mathbf{Sch}(L_{1})$, there is only one generator $L_{1}$. Thus, it is determined by the parameters $\xi_{1}, \eta_{1}, k_{1}$. However, $\xi_{1}, \eta_{1}$ are fixed by the normalized conditions. Therefore, the genus one Schottky group is completely determined by the multiplier $k_{1}$.

Different sets of parameters $(k_{1}, \eta_{2}, k_{2}, \xi_{3}, \eta_{3}, k_{3}, ..., \xi_{g}, \eta_{g}, k_{g})$ correspond to various Schottky groups. All genus $g$ ($g\geq 2$) Schottky groups form the genus $g$ Schottky space. For genus $g\geq 2$, the vector $(k_{1}, \eta_{2}, k_{2}, \xi_{3}, \eta_{3}, k_{3}, ..., \xi_{g}, \eta_{g}, k_{g})$ represents a point in the space $\mathbb{C}^{3g-3}$ ($\mathbb{C}$ is the complex plane). Thus, the dimension of the genus $g$ ($g\geq 2$) Schottky space is $6g-6$. Similarly, the dimension of the genus one Schottky space is two. Therefore, the dimension of the genus $g$ Schottky space equals the dimension of the genus $g$ Teichmüller space, which consists of the conformal equivalence classes of marked genus $g$ Riemann surfaces. Furthermore, the Teichm\"{u}ller space is the universal cover of the Schottky space.

We use $\mathbf{\Sigma}_{g}$ to denote the genus $g$ closed  Riemann surface. Then $\mathbf{\Sigma}_{g}$  can be constructed by the operation of the Schottky group on the Riemann sphere $\mathbb{S}$~\cite{KK}:
\begin{equation}
\label{eq:2.8}
\mathbf{\Sigma}_{g}=(\mathbb{S}-\mathbf{Fix})/ \mathbf{Sch}(L_{1}, L_{2}, ..., L_{g} ).
\end{equation}
Here, $\mathbf{Fix}$ denotes the set of fixed points of the Schottky group.  A fixed point of an element $L$ in the Schottky group is a point $z_{0}\in  \mathbb{S}$ that satisfies $L(z_{0})=z_{0}$. The term $\mathbb{S}- \mathbf{Fix}$ represents the deletion of the fixed points from the Riemann sphere $\mathbb{S}$. Equation \eqref{eq:2.8} represents  the Schottky parameterization of the Riemann surface $\mathbf{\Sigma}_{g}$. The Schottky parameterization of $\mathbf{\Sigma}_{g}$ models $\mathbf{\Sigma}_{g}$ as the quotient of $(\mathbb{S}- \mathbf{Fix})$ by the Schottky group $\mathbf{Sch}(L_{1}, L_{2}, ..., L_{g} )$.

Equation \eqref{eq:2.8} can be roughly interpreted in an intuitive way~\cite{KK}.  On the Riemann sphere $\mathbb{S}$, one can  choose $2g$ non-intersecting circles $C_{1}$, $ C_{1}^{'}$, $C_{2}$, $ C_{2}^{'}$, ..., $C_{g}$, $ C_{g}^{'}$ such that each circle lies  entirely outside the others. These circles are known as the Schottky circles.  Cut out the interior of each circle. Finally,  glue each pair of circles $C_{i}$ and $ C_{i}^{'}$ together to form a closed surface. The resulting surface is the Riemann surface $\mathbf{\Sigma}_{g}$.

Equation \eqref{eq:2.8} shows that a Schottky group corresponds to a Riemann surface $\mathbf{\Sigma}_{g}$. Thus, every point in the Schottky space corresponds to a Riemann surface $\mathbf{\Sigma}_{g}$. Therefore, the sets of parameters $(k_{1}, \eta_{2}, k_{2}, \xi_{3}, \eta_{3}, k_{3}, ..., \xi_{g}, \eta_{g}, k_{g})$ can be used to distinguish different Riemann surfaces $\mathbf{\Sigma}_{g}$. The AdS$_{3}$ handlebody can be obtained by ``filling'' the interior of the Riemann surface $\mathbf{\Sigma}_{g}$. Therefore, the Schottky space can describe various AdS$_{3}$ handlebodies.

The Schottky space is useful for representing the regularized action of the AdS$_{3}$ handlebody. However, it is less convenient for summing over physically non-equivalent AdS$_{3}$ handlebody spacetimes. This limitation arises because different points in the Schottky space may correspond to the same Riemann surface, and  the Schottky space is not simply connected~\cite{KK}. In our model, the Siegel upper half space ($\mathbf{Sie}(g)$) offers a more convenient framework for summing over different AdS$_{3}$ handlebodies.

\section{Relationship between the period matrix and the Schottky group }
\label{sec:A1}

The period matrix can also be used to distinguish between different Riemann surfaces $\mathbf{\Sigma}_{g}$. It belongs to the elements of the space $\mathbf{Sie}(g)$. The relationship between the period matrix and the Schottky group facilitates the transformation of functions from the Schottky space to the  space $\mathbf{Sie}(g)$. The space $\mathbf{Sie}(g)$ is  convenient for defining the integral domain when summing  over different AdS$_{3}$ handlebodies.

\begin{figure}[tbp]
\centering
\includegraphics[width=5cm]{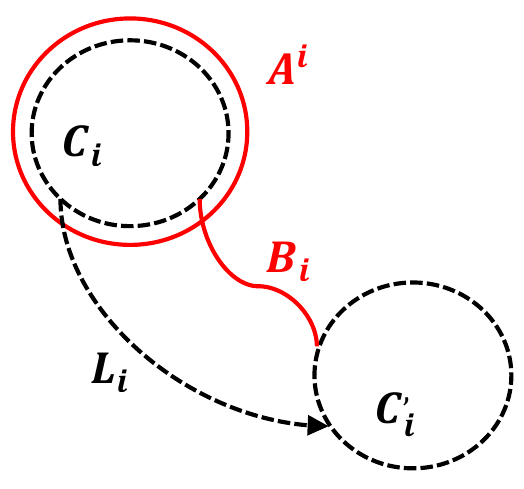}
\caption{\label{fig:1} Diagram of the Schottky circles and the canonical homology bases of circles. The black dashed circles represent the Schottky circles, while the red curves represent the canonical homology bases of circles. This figure is drawn on the Riemann sphere. The two ends of the curve $B_{i}$ become the same point after the operation of the Schottky group.}
\end{figure}

The generators of the homology group of the Riemann surface $\mathbf{\Sigma}_{g}$ are denoted as $\{A^{1}, A^{2}, ..., A^{g}; B_{1}, B_{2}, ..., B_{g}\}$. All of them are non-contractible circles on $\mathbf{\Sigma}_{g}$. Figure~\ref{fig:1} depicts the relationship between the circles $A^{i}$, $B_{i}$ and $C_{i}$, $C_{i}^{'}$. A similar figure can be found in references~\cite{BM,SP}.  The period matrix is defined as~\cite{M,EAV}
\begin{equation}
\label{eq:2.9}
\Pi(g)_{ij}\equiv \oint_{B_{j}}\omega_{i},
\end{equation}
where $\omega_{i}$ is defined by~\cite{BM,PFM}
\begin{equation}
\label{eq:2.10}
\omega_{i}\equiv \sum_{\mathscr{T}\in\mathbf{Sch}|L_{i}} (\frac{1}{z-\mathscr{T}(\eta_{i})}-\frac{1}{z-\mathscr{T}(\xi_{i})})dz.
\end{equation}
Here, the summation $\sum\limits_{\mathscr{T}\in\mathbf{Sch}|L_{i}}$ is over all elements $\mathscr{T}$ of the Schottky group $\mathbf{Sch}(L_{1}, L_{2}, ..., L_{g} )$, excluding those with $L_{i}^{n}$ ($n=\pm 1, \pm 2, ...$) as their rightmost factor. $\omega_{i}$ is a holomorphic differential on the Riemann surface $\mathbf{\Sigma}_{g}$. It is normalized along the $A^{i}$ circle, that is, $ \oint_{A^{j}}\omega_{i}=2\pi i \delta_{ij}$.

Substituting equation \eqref{eq:2.10} into \eqref{eq:2.9}, and completing the integration along the $B_{j}$ circle, one obtains~\cite{BM,PFM,LSR}
\begin{equation}
\label{eq:2.11}
2\pi i\Pi(g)_{ij}=\delta_{ij}\mathrm{ln}(k_{i})+\sum_{\mathscr{T}\in L_{i}|\mathbf{Sch}|L_{j}}\mathrm{ln}\big(\frac{(\eta_{i}-\mathscr{T}(\eta_{j}))(\xi_{i}-\mathscr{T}(\xi_{j}))}{(\eta_{i}-\mathscr{T}(\xi_{j}))(\xi_{i}-\mathscr{T}(\eta_{j}))}\big),
\end{equation}
where the summation $\sum\limits_{\mathscr{T}\in L_{i}|\mathbf{Sch}|L_{j}}$ is over all elements $\mathscr{T}$ of the Schottky group $\mathbf{Sch}(L_{1}, L_{2}, ..., L_{g} )$, excluding those with  $L_{i}^{n}$ ($n=\pm 1, \pm 2, ...$) as the leftmost factor or $L_{j}^{n}$ as the rightmost factor. In the case where $i=j$, the identity element of $\mathbf{Sch}(L_{1}, L_{2}, ..., L_{g} )$ is also excluded from  the summation.

Equation \eqref{eq:2.11} establishes the relationship between the period matrix and the Schottky group. However, it is difficult to calculate the summation on the right hand side of equation \eqref{eq:2.11} strictly. A first order approximation \eqref{eq:2.6} of the Schottky group $\mathbf{Sch}(L_{1}, L_{2}, ..., L_{g} )$ suffices for our model (see appendix~\ref{sec:A0}).  Under this approximation, only the elements $\{L_{j}, L_{j}^{-1} \ | j=1, 2, ..., g;\ j\neq i\}$ contribute to the summation in equation \eqref{eq:2.11} when calculating the element $\Pi(g)_{ii}$. For the calculation of the non-diagonal element $\Pi(g)_{ij}$ ($i\neq j$), one only needs to sum over the elements $\{id, L_{n}, L_{n}^{-1} \ | n=1, 2, ..., g;\ n\neq i; \ n\neq j\}$.

Using this approximation,  the period matrix can be simplified to
\begin{equation}
\label{eq:2.12}
2\pi i \Pi(g)_{ii}=\mathrm{ln}(k_{i})+\sum_{n\neq i}\frac{2k_{n}(\xi_{i}-\eta_{i})^{2}(\xi_{n}-\eta_{n})^{2}}{(\xi_{i}-\xi_{n})(\eta_{i}-\eta_{n})(\eta_{i}-\xi_{n})(\xi_{i}-\eta_{n})} +\sum_{n \neq i}o(k_{n}^{2}),
\end{equation}
\begin{equation}
\label{eq:2.13}
2\pi i \Pi(g)_{ij}=\mathrm{ln}\frac{(\eta_{i}-\eta_{j})(\xi_{i}-\xi_{j})}{(\eta_{i}-\xi_{j})(\xi_{i}-\eta_{j})}+\sum_{n\neq i,j}\frac{2k_{n}(\xi_{n}-\eta_{n})^{2}(\eta_{i}-\xi_{i})(\eta_{j}-\xi_{j})}{(\eta_{i}-\xi_{n})(\xi_{i}-\xi_{n})(\eta_{j}-\eta_{n})(\xi_{j}-\eta_{n})}+\sum_{n \neq i,j}o(k_{n}^{2}).
\end{equation}
In equation \eqref{eq:2.12}, $n=1, 2, ..., g$ and $n\neq i$. In equation \eqref{eq:2.13}, $n=1, 2, ..., g$ and $n\neq i, j$.     To derive equations  \eqref{eq:2.12} and \eqref{eq:2.13} from \eqref{eq:2.11}, we used the formula~\cite{BM}
\begin{equation}
\label{eq:2.14}
\frac{(z_{1}-L_{i}(\sigma_{1}))(z_{2}-L_{i}(\sigma_{2}))}{(z_{1}-L_{i}(\sigma_{2}))(z_{2}-L_{i}(\sigma_{1}))}=1+\frac{k_{i}(\xi_{i}-\eta_{i})^{2}(z_{1}-z_{2})(\sigma_{1}-\sigma_{2})}{(z_{1}-\xi_{i})(z_{2}-\xi_{i})(\sigma_{1}-\eta_{i})(\sigma_{2}-\eta_{i})}+o(k_{i}^{2}).
\end{equation}
Here, $z_{1}$, $z_{2}$, $\sigma_{1}$, $\sigma_{2}$ are points on the Riemann surface $\mathbf{\Sigma}_{g}$. In the case where the Schottky group contains only two generators, equations \eqref{eq:2.12} and \eqref{eq:2.13} reduce to equation (A.28) in reference~\cite{BM}.

\begin{figure}[tbp]
\centering
\includegraphics[width=14cm]{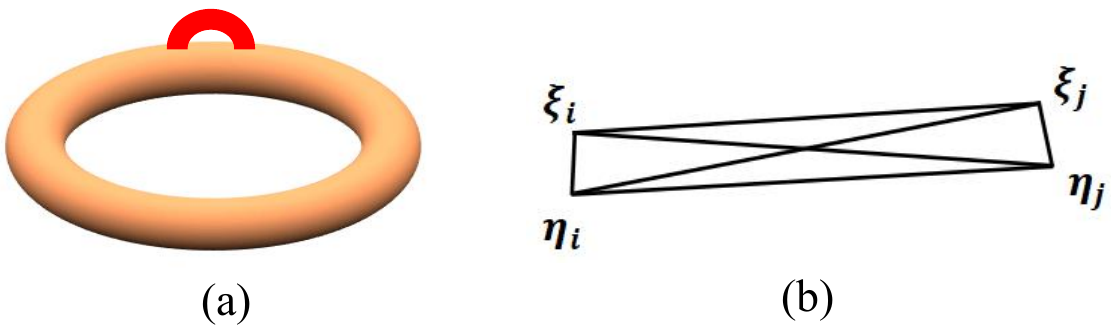}
\caption{\label{fig:2} A diagram illustrating boundary wormhole fluctuations and the dilute wormhole gas approximation. Figure (a) depicts the boundary wormhole fluctuations, while figure (b) illustrates the dilute wormhole gas approximation. In figure (a), the orange torus represents the conformal boundary of a Euclidean thermal AdS$_{3}$ spacetime, and the red strip represents a wormhole with two ends.  In figure (b), $\xi_{i}$ and $\eta_{i}$ approximately denote the two ends of one wormhole, while $\xi_{j}$ and $\eta_{j}$ represent the two ends of another wormhole. }
\end{figure}

Different AdS$_{3}$ handlebody spacetimes can be characterized by the period matrix. For instance, the conformal boundary of the Euclidean  thermal AdS$_{3}$ spacetime (TAdS$_{3}$) is a torus. The period matrix of a torus is a complex number, which is the modular parameter of the torus. The real part of the period matrix for TAdS$_{3}$ corresponds to the angular potential, while its imaginary part corresponds to the inverse of temperature. The time circle in TAdS$_{3}$ is non-contractible. Figure~\ref{fig:2}(a) illustrates the creation of a wormhole in TAdS$_{3}$.  The resulting spacetime is a genus two handlebody, where the period matrix is a $2\times2$ symmetric matrix.

Equations \eqref{eq:2.12} and \eqref{eq:2.13} are still complicated. We need to simplify them further. To achieve this,  we consider the dilute wormhole gas approximation~\cite{SK,AS}, which implies that the size of the wormholes is much smaller than their distance. Note that under the small $k_{i}$ limit, $\xi_{i}$ and $\eta_{i}$ approximately represent the two ends of the wormhole~\cite{AS}. Thus, in this case, one can use the equation
\begin{equation}
\label{eq:2.15}
|\xi_{i}-\eta_{i}|\ll |\xi_{i}-\eta_{j}| \quad \quad (i\neq j)
\end{equation}
to represent the dilute wormhole gas approximation.  Figure~\ref{fig:2}(b) depicts the dilute wormhole gas approximation visually. It also illustrates that the differences between $|\xi_{i}-\eta_{j}|$, $|\xi_{i}-\xi_{j}|$, and $|\eta_{i}-\eta_{j}|$ $(i\neq j)$ are small.

Using equation \eqref{eq:2.15}, the period matrix in equations \eqref{eq:2.12} and \eqref{eq:2.13} can be approximated as
\begin{equation}
\label{eq:2.16}
2\pi i \Pi(g)_{ii}\approx\mathrm{ln}(k_{i}),
\end{equation}
\begin{equation}
\label{eq:2.17}
2\pi i \Pi(g)_{ij}\approx\mathrm{ln}\frac{(\eta_{i}-\eta_{j})(\xi_{i}-\xi_{j})}{(\eta_{i}-\xi_{j})(\xi_{i}-\eta_{j})}.
\end{equation}
Equations \eqref{eq:2.16} and \eqref{eq:2.17} describe the relationship between the period matrix and the Schottky group under the dilute wormhole gas approximation.
These two equations are consistent with formula (3.3) in reference~\cite{AS}. In addition, when deriving equations \eqref{eq:2.16} and \eqref{eq:2.17}, we have used the small $k_{j}$ limit. In this limit, $1-\mathrm{exp}(2\pi i \Pi(g)_{jj})\approx 1$ (this approximation had been used by Lyons and Hawking in~\cite{AS}).

Following the discussions by Lyons and Hawking in reference~\cite{AS}, if the generators $L_{i}$ and $L_{j}$ correspond to different wormholes,  the element $\Pi(g)_{ij}$ ($i\neq j$)  can be interpreted as the interaction between wormholes $i$ and  $j$. Thus, equation \eqref{eq:2.17} indicates that the interactions between different wormholes are small under the dilute wormhole gas approximation. In a special case where all non-diagonal elements of the period matrix are zero, $\Pi(g)=\Pi(g)_{11}\oplus\Pi(g)_{22}\oplus\cdot\cdot\cdot\oplus\Pi(g)_{gg}$,  representing a set of disconnected tori, with the modular parameter of the  $i$-th torus being $\Pi(g)_{ii}$.

The Siegel upper half space $\mathbf{Sie}(g)$ consists of all $g\times g$ symmetric matrices where the imaginary part of each matrix element is positive~\cite{HM}. Formally, the space $\mathbf{Sie}(g)$ is defined as $\mathbf{Sie}(g)\equiv\{\Omega \ |\ \Omega_{ij}=\Omega_{ji},\ \mathrm{Im}(\Omega_{ij})>0\}$. The $g\times g$ period matrix is also symmetric ($\Pi(g)_{ij}=\Pi(g)_{ji}$) and belongs to the elements of the space $\mathbf{Sie}(g)$. Equations \eqref{eq:2.16} and \eqref{eq:2.17} facilitate transforming the tunneling action from the Schottky space to the space $\mathbf{Sie}(g)$. This helps in summing over different AdS$_{3}$ handlebodies.

\section{Derivation for the vacuum decay seed $B_{seed}$  }
\label{sec:A}

In the region $\rho<\rho_{s}$, it can be shown that equation \eqref{eq:3.12} reduces to
\begin{eqnarray}\begin{split}
\label{eq:A1}
S_{E}(\rho\leq\rho_{s})=8\pi\int_{0}^{\rho_{s}}d\tau_{E}\big\{-1+\rho^{2}V(\phi)\big\}.
\end{split}
\end{eqnarray}
Inside the bubble, $\phi_{bounce}=\phi_{1}$, $\dot{\phi}=0$ and $d\rho=\sqrt{1-\rho^{2}V(\phi_{1})}d\tau_{E}$. This relationship can be derived from equation \eqref{eq:3.6} by setting $\dot{\phi}=0$. Combining equations \eqref{eq:3.8} and \eqref{eq:A1} with these conclusions,  the tunneling action $B_{in}$ becomes
\begin{equation}
\label{eq:3.9A}
B_{in}=-8\pi\int_{0}^{\bar{\rho}}d\rho[1-\rho^{2}V(\phi_{1})]^{\frac{1}{2}}+8\pi\int_{0}^{\bar{\rho}}d\rho[1-\rho^{2}V(\phi_{2})]^{\frac{1}{2}}.
\end{equation}
Using the integral formula
\begin{equation}
\label{eq:3.9aA}
\int_{0}^{a}\sqrt{1-bx^{2}}dx=\frac{1}{2}\big\{a\sqrt{1-a^{2}b}+b^{-\frac{1}{2}}\ \mathrm{arcsin}(a\sqrt{b})\big\},
\end{equation}
and performing the integral over the variable $\rho$, then equation \eqref{eq:3.9A} becomes
\begin{eqnarray}\begin{split}
\label{eq:3.9bA}
B_{in}=&4\pi\big\{V^{-\frac{1}{2}}(\phi_{2})\ \mathrm{arcsin}(\bar{\rho}\sqrt{V(\phi_{2})})-V^{-\frac{1}{2}}(\phi_{1})\ \mathrm{arcsin}(\bar{\rho}\sqrt{V(\phi_{1})})\big\}\\&+4\pi\bar{\rho}\big\{\sqrt{1-\bar{\rho}^{2}V(\phi_{2})}-\sqrt{1-\bar{\rho}^{2}V(\phi_{1})}\big\}.
\end{split}
\end{eqnarray}
Equation \eqref{eq:3.9bA} represents the tunneling action inside the bubble.

The tension ($\mu$) of the domain wall is given by $\mu=\int_{\phi_{1}}^{\phi_{2}}d\phi\sqrt{2(V(\phi)-V(\phi_{2}))}$. Under the thin wall approximation, the tension $\mu$  and the tunneling action $B_{wall}$ are equal to the energy density and the total energy of the domain wall, respectively~\cite{SEA}.  Therefore, the tunneling action $B_{wall}$ is
\begin{equation}
\label{eq:3.10A}
B_{wall}=4\pi\bar{\rho}^{2}\mu=4\pi\bar{\rho}^{2}\int_{\phi_{1}}^{\phi_{2}}d\phi\sqrt{2(V(\phi)-V(\phi_{2}))}.
\end{equation}
Equation \eqref{eq:3.10A} is the lower dimensional version of equation (2.15) in reference~\cite{SF}. A more detailed explanation of this method for calculating the tunneling action can be found in~\cite{SF}.

From equations \eqref{eq:3.11}, \eqref{eq:3.9bA} and \eqref{eq:3.10A},   the vacuum decay seed $B_{seed}$ can be expressed as
\begin{eqnarray}\begin{split}
\label{eq:3.27A}
B_{seed}=4\pi\Big\{&V^{-\frac{1}{2}}(\phi_{2})\ \mathrm{arcsin}(\bar{\rho}\sqrt{V(\phi_{2})})-V^{-\frac{1}{2}}(\phi_{1})\ \mathrm{arcsin}(\bar{\rho}\sqrt{V(\phi_{1})})\\&+\bar{\rho}\sqrt{1-\bar{\rho}^{2}V(\phi_{2})}-\bar{\rho}\sqrt{1-\bar{\rho}^{2}V(\phi_{1})}+\bar{\rho}^{2}\mu\Big\}.
\end{split}
\end{eqnarray}
By taking $\partial B/\partial \bar{\rho}=0$,  the bubble radius $\bar{\rho}$ can be obtained as
\begin{equation}
\label{eq:3.28A}
\bar{\rho}=\frac{2\mu}{\sqrt{\mu^{4}+\epsilon^{2}+2\mu^{2}(2V(\phi_{2})+\epsilon)}}.
\end{equation}
Equation \eqref{eq:3.28A} indicates that the radius $\bar{\rho}$ is determined by the tension of the domain wall and the energy density of the vacuum states.

\section{Derivation for equation \eqref{eq:3.44} }
\label{sec:B}

The tunneling action $B_{out}$ in equation \eqref{eq:3.36} is complicated, so we need to simplify it. Defining $\mathscr{E}_{j}\equiv\xi_{j}-\eta_{j}$, then equation \eqref{eq:2.17} implies that the element $\Pi(g)_{2j}$ can be expressed as
\begin{equation}
\label{eq:3.37}
\Pi(g)_{2j}=\frac{1}{2\pi i}\mathrm{ln}\big(1+\frac{\mathscr{E}_{j}}{\eta_{2}-\xi_{j}}+\frac{\mathscr{E}_{j}}{\eta_{j}-\xi_{2}}-\frac{\mathscr{E}_{j}^{2}}{(\eta_{2}-\xi_{j})(\xi_{2}-\eta_{j})}\big).
\end{equation}
Under the dilute wormhole gas approximation, where the size of the wormholes is significantly smaller than the distance between them (equation \eqref{eq:2.15}). Thus, equation \eqref{eq:3.37} can be approximated as
\begin{equation}
\label{eq:3.38}
\Pi(g)_{2j}\approx\frac{1}{2\pi i}(\frac{\mathscr{E}_{j}}{\eta_{2}-\xi_{j}}+\frac{\mathscr{E}_{j}}{\eta_{j}-\xi_{2}}).
\end{equation}
Equation \eqref{eq:3.38} indicates that under the dilute wormhole gas approximation, the element $\Pi(g)_{2j}$ is small.

However, the element $\Pi(g)_{1j}$ represents the interaction between the parent universe and the wormhole. The dilute wormhole gas approximation does not imply that $\Pi(g)_{1j}$ is small.  Therefore, from equations \eqref{eq:3.33}-\eqref{eq:3.35}, the quantities $\Delta_{1j}$, $\Delta_{2j}$, and $\Delta_{3j}$ can be approximated as:
\begin{equation}
\label{eq:3.39}
\Delta_{1j}\approx 1-\mathrm{exp}(2\pi i \Pi(g)_{1j}),\ \ \ \ \ \ \ \ \ \ \ \ \ \Delta_{3j}\approx 0,
\end{equation}
\begin{equation}
\label{eq:3.40}
\Delta_{2j}\approx \big(1-\mathrm{exp}(2\pi i \Pi(g)_{1j})\big)\big(1-\mathrm{exp}(-2\pi i \Pi(g)_{12})\big).
\end{equation}
Equations \eqref{eq:3.39} and \eqref{eq:3.40} imply that the interactions between different wormholes are neglected, while the interactions between the parent universe and the wormholes are retained.

Substituting equations \eqref{eq:3.39} and \eqref{eq:3.40} into equation \eqref{eq:3.36}, one obtains $B_{out}=\infty$ or
\begin{eqnarray}\begin{split}
\label{eq:3.41}
B_{out}=\frac{-4\pi}{\sqrt{-V(\phi_{2})}}\Big\{&\sum_{j=2}^{g}\mathrm{ln}\Big|\frac{\mathrm{exp}(\pi i \Pi(g)_{jj})(\mathrm{exp}(2 \pi i \Pi(g)_{1j})-1)}{1-\mathrm{exp}(2\pi i \Pi(g)_{jj})}\Big|\\&+\sum_{j=2}^{g}\mathrm{ln}\Big|(\mathrm{exp}(-2\pi i \Pi(g)_{12})-1)\Big|\Big\}.
\end{split}
\end{eqnarray}
The symbol $``\pm"$ in equation \eqref{eq:3.36} corresponds to two different tunneling actions $B_{out}$.  The case $B_{out}=\infty$ ( associated with $``+"$ )is unphysical and is therefore disregarded.  Equation \eqref{eq:3.41} corresponds to the $``-"$ case.

The period matrix can be written as $\Pi(g)=X+iY$. Then, it is straightforward to prove that
\begin{eqnarray}\begin{split}
\label{eq:3.42}
&\ \ \ \Big|\frac{\mathrm{exp}(\pi i \Pi(g)_{jj})(\mathrm{exp}(2 \pi i \Pi(g)_{1j})-1)}{1-\mathrm{exp}(2\pi i \Pi(g)_{jj})}(\mathrm{exp}(-2\pi i \Pi(g)_{12})-1)\Big|\\&=\mathrm{exp}(-\pi Y_{jj})\cdot \Big(\mathrm{exp}(-4\pi Y_{jj})-2 \mathrm{cos}(2\pi X_{jj})\mathrm{exp}(-2\pi Y_{jj})+1\Big)^{-\frac{1}{2}}\\&\ \ \  \times \Big(\mathrm{exp}(-4\pi Y_{1j})-2 \mathrm{cos}(2\pi X_{1j})\mathrm{exp}(-2\pi Y_{1j})+1\Big)^{\frac{1}{2}}\\&\ \ \ \times\Big(\mathrm{exp}(4\pi Y_{12})-2 \mathrm{cos}(2\pi X_{12})\mathrm{exp}(2\pi Y_{12})+1\Big)^{\frac{1}{2}}.
\end{split}
\end{eqnarray}
Combining equations \eqref{eq:3.43}, \eqref{eq:3.41} and \eqref{eq:3.42}, the tunneling action $B_{out}$ can be expressed as
\begin{eqnarray}\begin{split}
\label{eq:3.44a}
B_{out}=&-\Xi(\phi_{2})(g-1)\mathrm{ln}\Big(1+\mathrm{exp}(4\pi Y_{12})-2\mathrm{cos}(2\pi X_{12})\mathrm{exp}(2\pi Y_{12})\Big)\\&-\Xi(\phi_{2})\sum_{j=2}^{g}\Big\{-2\pi Y_{jj}-\mathrm{ln}\Big(1+\mathrm{exp}(-4\pi Y_{jj})-2\mathrm{cos}(2\pi X_{jj})\mathrm{exp}(-2\pi Y_{jj})\Big)\\&\ \ \ \ \ \ \ \ \    +\mathrm{ln}\Big(1+\mathrm{exp}(-4\pi Y_{1j})-2\mathrm{cos}(2\pi X_{1j})\mathrm{exp}(-2\pi Y_{1j})\Big)\Big\}.
\end{split}
\end{eqnarray}

\end{document}